\documentclass[10pt,letterpaper,conference]{IEEEtran}
\usepackage{booktabs} % For formal tables
\usepackage[flushleft]{threeparttable}
\usepackage{xcolor}
\usepackage[utf8]{inputenc}
\usepackage{float}
\usepackage{amssymb}
\usepackage{amsmath}
\usepackage{ifthen}
\usepackage{caption}
\usepackage[bookmarks=false]{hyperref}
\usepackage{url,moreverb,xspace}
\usepackage{enumitem}
\usepackage{array,graphicx}
\usepackage{soul}
\usepackage{balance}
\usepackage{pifont}
\usepackage{nicefrac}
\usepackage{mathtools}
\usepackage{tabularx}
\usepackage{xcolor,pifont}
\usepackage{tikz}
\usepackage{svg}
\usepackage{pdfpages}
\usepackage{array}
\usepackage{subcaption}
\usepackage[most]{tcolorbox}
\usepackage{multicol}
\usepackage{multirow}
\usepackage{pgfplots}
\usepackage{xcolor}
\usetikzlibrary{patterns}
\pgfplotsset{compat=1.18}
\usepackage{makecell}
\usepackage[normalem]{ulem}

\newcommand*\colourcheck[1]{%
  \expandafter\newcommand\csname #1check\endcsname{\textcolor{#1}{\ding{52}}}%
}
\newcommand*\colourcross[1]{%
  \expandafter\newcommand\csname #1cross\endcsname{\textcolor{#1}{\ding{56}}}%
}
\usepackage[vlined, boxruled, linesnumbered] {algorithm2e}
\SetKw{KwBy}{by}
\SetKw{KwBreak}{break}
\SetKw{KwReturn}{return}
\def\HiLi{\leavevmode\rlap{\hbox to \hsize{\color{gray!35}\leaders\hrule height .8\baselineskip depth .5ex\hfill}}}

\usepackage{amsthm}

\theoremstyle{definition} % Non-italic, upright text
\newtheorem{definition}{Definition}[section] % 
\newtheorem{example}{Example}[section] % 

\newlength\BARWIDTH
\newlength\BARHEIGHT
\SetKwComment{Comment}{$\triangleright$\ }{}
\SetCommentSty{itshape}
\newboolean{showcomments}
\setboolean{showcomments}{true}
\ifthenelse{\boolean{showcomments}}
{\newcommand{\nb}[2] {
  \fcolorbox{black}{gray!20}{\bfseries\sffamily\scriptsize#1:}
  {\sf\small$\blacktriangleright$\textit{#2}$\blacktriangleleft$}
}
}
{\newcommand{\nb}[2]{}
}

\newcommand{\head}[1]{\noindent\textbf{#1.}}

\newcounter{fcounter}
\makeatletter
\newcommand{\thickhline}{%
    \noalign {\ifnum 0=`}\fi \hrule height 1pt
    \futurelet \reserved@a \@xhline
}
\makeatother
\usepackage{fontawesome5}

\newcommand{\appname}{STELLAR\xspace} % 
 
\newcommand{\rs}{\textsc{RS}\xspace}
\newcommand{\astral}{\textsc{ASTRAL}\xspace}
\newcommand{\gs}{\textsc{T-wise}\xspace}

\newcommand{\deepseekcloud}{\textsc{DeepSeek-V3}\xspace}
\newcommand{\deepseeklocal}{\textsc{DeepSeek-V2-16B}\xspace}

\newcommand{\gptfouro}{\textsc{GPT-4o}\xspace}

\newcommand{\gptfivechat}{\textsc{GPT-5-Chat}\xspace}
\newcommand{\gptthreefive}{\textsc{GPT-3.5}\xspace}

\newcommand{\llama}{\textsc{Llama~3}\xspace}
\newcommand{\qwentwo}{\textsc{Qwen~2.5-7B}\xspace}
\newcommand{\qwenthree}{\textsc{Qwen~3-8B}\xspace}

\newcommand{\mistral}{\textsc{Mistral~7B}\xspace}
\newcommand{\dolphinthree}{\textsc{Dolphin~3}\xspace}
\newcommand{\gptfouromini}{\textsc{GPT-4o-Mini}\xspace}

\newcommand{\llamathreetwo}{\textsc{Llama~3.2}\xspace}

\usepackage{wrapfig}

\newcommand{\company}{\textsc{BMW}\xspace}
\newcommand{\convnavione}{NaviQA-I\xspace}
\newcommand{\convnavitwo}{NaviQA-II\xspace}

\usepackage{verbatim}
\usepackage{minted}

\usepackage{tablefootnote}

\definecolor{jsonkey}{rgb}{0.0,0.0,0.6}      % dark blue for keys
\definecolor{jsonstring}{rgb}{0.6,0.0,0.0}   % dark red for string values
\definecolor{jsonnumber}{rgb}{0.0,0.5,0.0}   % dark green for numbers
\definecolor{jsonpunct}{rgb}{0.13,0.13,0.13} % punctuation
\definecolor{jsonbg}{rgb}{0.95,0.95,0.95}    % background

\lstdefinelanguage{JSON}{
    basicstyle=\ttfamily\footnotesize,
    breaklines=true,
    frame=single,
    backgroundcolor=\color{jsonbg},
    literate=
    *{"}{{{\color{jsonkey}"}}}{1}
    {:}{{{\color{jsonpunct}{:}}}}{1}
    {,}{{{\color{jsonpunct}{,}}}}{1}
    {\{}{{{\color{jsonpunct}{\{}}}}{1}
    {\}}{{{\color{jsonpunct}{\}}}}}{1}
    {[}{{{\color{jsonpunct}{[}}}}{1}
    {]}{{{\color{jsonpunct}{]}}}}{1}
}

\begin{document}

\pagenumbering{arabic} 
\pagestyle{plain}

%\title{Automated Testing of In-Car Conversational Systems using Large Language Models}

\title{
\appname: A Search-Based Testing Framework for Large Language Model Applications}

\author{
    Lev Sorokin, Ivan Vasilev, Ken E. Friedl \\
    BMW Group \\
    \{lev.sorokin, ivan.vasilev, ken.friedl\}@bmw.de
    \and
    Andrea Stocco \\
    Technical University of Munich, fortiss GmbH \\
    andrea.stocco@tum.de, stocco@fortiss.org
}

%\renewcommand{\shortauthors}{Giebisch et al.}

% \IEEEoverridecommandlockouts
% \IEEEpubid{\makebox[\columnwidth]{XXX/\$31.00~\copyright2024 IEEE \hfill} \hspace{\columnsep}\makebox[\columnwidth]{ }}

\maketitle

\begin{abstract}
Large Language Model (LLM)-based applications are increasingly deployed across various domains, including customer service, education, and mobility. However, these systems are prone to inaccurate, fictitious, or harmful responses, and their vast, high-dimensional input space makes systematic testing particularly challenging. To address this, we present \appname, an automated search-based testing framework for LLM-based applications that systematically uncovers text inputs leading to inappropriate system responses. Our framework models test generation as an optimization problem and discretizes the input space into stylistic, content-related, and perturbation features. 
Unlike prior work that focuses on prompt optimization or coverage heuristics, our work employs evolutionary optimization to dynamically explore feature combinations that are more likely to expose failures. 
We evaluate \appname on three LLM-based conversational question-answering systems. The first focuses on safety, benchmarking both public and proprietary LLMs against malicious or unsafe prompts. The second and third target navigation, using an open-source and an industrial retrieval-augmented system for in-vehicle venue recommendations. Overall, \appname exposes up to 4.3$\times$ (average 2.5$\times$) more failures than the existing baseline approaches.
\end{abstract}

\begin{IEEEkeywords}
large language models, LLM testing, in-car conversational systems, search-based testing
\end{IEEEkeywords}

% !TEX root =  paper.tex
\section{Introduction}\label{sec:introduction}

Large language models (LLMs) are employed in various domains, including text summarization~\cite{zhang2025comprehensivesurveyprocessorientedautomatic}, translation~\cite{elshin-etal-2024-general}, education~\cite{duolingo2024max}, coding tasks~\cite{dong2025surveycodegenerationllmbased,jin2025llmsllmbasedagentssoftware,liu2024largelanguagemodelbasedagents,10.1145/3769082}, or in more complex systems such as conversational assistance systems~\cite{friedl2023incarethinkingincarconversational,2025-Giebisch-IV, ahmed2025rag, rapisarda2025qatesting, bmw2024ces} and autonomous driving~\cite{2025-Guo-arxiv}. 
% Especially the automotive domain is exploring its usage in navigational assistant systems to improve the user experience and increase safety by enabling seamless voice-based interactions. 
% \item Task-oriented conversational systems
% \item Robustness important
However, a major challenge in integrating LLMs into software systems is their complex black-box nature, which can generate incorrect, incomplete, or even harmful information (e.g., hallucinations), such as providing users with recommendations on how to commit crimes. Therefore, comprehensive testing of LLM-based applications is essential to ensure their robustness.

Static large benchmark datasets~\cite{xie2025sorrybench, zang2020multiwoz,ji2023beavertails} are commonly used to evaluate the performance of LLMs across diverse inputs. However, they face two major limitations. First, data contamination undermines the trustworthiness of validation scores, as benchmark samples may inadvertently be included in the training data of the LLM~\cite{bradbury2025combinatorial}. Second, LLM-based applications must handle unexpected and diverse natural language inputs, making it infeasible to generate a dataset that exhaustively covers all possible test cases.
Automated test generation approaches such as ASTRAL~\cite{ASTRAL} explore feature combinations via a coverage matrix, but their application is constrained given the combinatorial explosion as the number of features grows. For example, testing an LLM's robustness against malicious inputs across eight feature types (e.g., content, persuasion, misspelling ratio), each with at least five possible values, would result in over 390,000 combinations. Assuming an average generation and execution time of 5 seconds per test, this would require more than 20 days of continuous execution, making large-scale testing impractical.

One testing paradigm particularly well-suited to handling complex and multidimensional input spaces is search-based software testing~\cite{Zeller17SBST}, which frames the testing process as an optimization problem, guiding the systematic generation of failure-revealing test inputs~\cite{5954405,riccio2020model,2020-Riccio-EMSE,2025-Weissl-TOSEM,sbse2019ramirez,Abdessalem-ICSE18,Abdessalem-ASE18-1,sorokin2023opensbt,NejatiSSFMM23,sorokin24svm,2025-Chen-EMSE,2025-Maryam-ICST,stocco2019misbehaviour} to reduce the testing costs and time.
% SBST has since been applied across diverse domains, ranging from code~\cite{5954405}, deep learning software~\cite{riccio2020model,2020-Riccio-EMSE,2025-Weissl-TOSEM}, up to complex cyber-physical systems, such as automated driving systems and unmanned aerial vehicles~\cite{sbse2019ramirez, NejatiSSFMM23, Abdessalem-ICSE18, Abdessalem-ASE18-1, sorokin2023opensbt, NejatiSSFMM23, sorokin24svm}. SBST frames the testing process as an optimization problem, guiding the systematic generation of failure-revealing test inputs. Candidate inputs are iteratively evolved by applying modifications to existing ones and evaluated using a fitness function that quantifies their fault-revealing potential.
While optimization algorithms have already been applied in the context of LLMs for prompt optimization~\cite{evoprompt}, they have not yet been systematically explored for the testing of LLM-based systems.
% \ken{add that SBT specifically target the challenge of cost/time disdvantage of large datasets. They make testing of multi dimensional systems possible at all given limited time and budget.}

% In this work, we introduce \appname, a search-based testing framework designed to systematically uncover text inputs that trigger inappropriate or faulty responses in LLM-based applications. The input space of natural language is discretized into stylistic, content-related, and perturbation features, enabling an evolutionary algorithm to efficiently explore combinations of characteristics and generate inputs more likely to reveal failures.
% The novelty of \appname lies in combining search-based testing with feature discretization. While prior work on LLM testing has largely focused on prompt optimization~\cite{evoprompt} or coverage heuristics~\cite{ASTRAL}, our approach provides a more systematic and scalable method for navigating the high-dimensional input space. This not only improves the effectiveness of failure detection by targeting diverse categories of problematic behaviors but also enhances the efficiency of the testing process by reducing redundant executions.
% As a result, \appname offers a generalizable strategy for benchmarking and stress-testing the robustness of LLM-based systems across different domains.
In this work, we introduce \appname, a search-based testing framework that systematically generates text inputs to expose inappropriate or faulty responses in LLM-based applications. By discretizing natural language into stylistic, content, and perturbation features, \appname uses evolutionary search to efficiently explore diverse input combinations. Unlike prior work focused on prompt tuning~\cite{evoprompt, pryzant-etal-2023-automatic} or coverage heuristics~\cite{ASTRAL}, our approach offers a scalable way to navigate high-dimensional input spaces, improving both the effectiveness of failure discovery and the efficiency of testing. 

We evaluate \appname on two use cases.
% \ken{make more clear what "domain" means in this context. Not only use cases, but also different system architectures.}.
The first, SafeQA, tests public and proprietary LLMs under malicious and safety-critical prompts. The second, NaviQA, involves two retrieval-augmented assistants in an in-car dialogue setting for navigational venue recommendations. Across both, \appname consistently reveals more failing inputs than random, combinatorial, and coverage-based baselines such as ASTRAL, demonstrating its effectiveness in assessing LLM robustness.

The contributions of this paper are as follows: 

\begin{itemize}  
    \item \textbf{Testing Framework.} \appname, an automated search-based testing framework that discretizes linguistic and semantic input features to systematically uncover failures in LLM-based applications, publicly available~\cite{repo}.

    \item \noindent \textbf{Evaluation.} An empirical study on two representative case studies that shows that \appname detects substantially more failing inputs than state-of-the-art automated testing techniques and random search.

    % \item \noindent \textbf{Replication.} We provide a replication package containing the implementation of \appname, and experiment scripts for both case studies to support reproducibility~\cite{repo}.
\end{itemize}

\section{Background and Motivation}\label{sec:background}

In the following, we provide the definitions and illustrative examples required to understand our work.

\begin{definition}
A large language model application is a software system that leverages a large language model to process semi-structured or unstructured inputs (e.g., natural language text) and generate corresponding outputs, typically in the form of semi-structured or unstructured text.
\end{definition}

We focus on text-based LLM applications, for which we designed a search-based testing approach defined as:

% \begin{definition}
% A search-based testing problem $P$ is defined as a tuple $P = (T, D, F, O)$, where

% \begin{itemize}
%     \item $T$ is the LLM-based system under test.
    
%     \item $D \subseteq \mathbb{R}^n$ is the search domain, where $n$ is the \textit{dimension} of the search space. The vector $\mathbf{x} =(x_1, \ldots, x_n) \in D$ is called test input. 
    
%     \item $F$ is the vector-valued fitness function defined as $F: D \mapsto \mathbb{R}^m, F(\mathbf{x}) = (f_1(\mathbf{x}),\ldots, f_m(\mathbf{x}))$, where $f_i$ is a scalar fitness function (or fitness function, for short) that assigns a quantitative value to each test input and $\mathbb{R}^m$ is the \textit{objective space}, and $m$ corresponds to the number of fitness functions. A fitness function evaluates how \textit{fit} a test input is, assigning a \emph{fitness value} to it.
    
%     \item $O$ is the oracle function, $O : \mathbb{R}^m \mapsto \lbrace 0,1 \rbrace $, which evaluates, given the objective space of the fitness functions, whether a test passes or fails. A test that fails is called \emph{failure-inducing}.
% \end{itemize}
% \end{definition}

\begin{definition}
A search-based testing problem for an LLM-based application $P$ is defined as a tuple $P = (AUT, D, F, O)$, where

\begin{itemize}
    \item $AUT$ is the LLM-based application under test.
    
    \item $D \subseteq \mathbb{R}^n$ is the search domain, where $n$ is the \textit{dimension} of the search space. The vector $\mathbf{x} =(x_1, \ldots, x_n) \in D$ represents a candidate textual input (test case) to the AUT. 
    
    \item $F$ is the vector-valued fitness function defined as $F: D \mapsto \mathbb{R}^m, \; F(\mathbf{x}) = (f_1(\mathbf{x}),\ldots, f_m(\mathbf{x}))$, where each $f_i$ is a scalar fitness function that assigns a quantitative score to an input based on its ability to expose faulty or undesired LLM behavior. The objective space $\mathbb{R}^m$ corresponds to the number of evaluation criteria. 
    
    \item $O$ is the oracle function, $O : \mathbb{R}^m \mapsto \{0,1\}$, which decides whether a generated input reveals a failure. An input for which $O(F(\mathbf{x}))=1$ is considered \emph{failure-inducing}.
\end{itemize}
\end{definition}

\begin{example}
% \ken{Why do we only test using textual in- and output? Arguments could be: easier to implement/deploy, reduction of error sources such as noise or dialets of speaking, etc. } 
An example of an LLM-based application is NaviQA, one of the experimental subjects used in this work. It is a task-oriented dialogue system used in commercial vehicles~\cite{bmw2024ces, rony-etal-2023-carexpert}, whose goal is to provide venue recommendations based on user utterances. %\hl{
While the actual user input is provided through voice-based commands, we focus here on text-based user input and system output, 
to mitigate the influence of error sources such as noise or background voices in performing the user requests and evaluating the outcome.%}

For instance, an input to the system is the following user utterance: \textit{``Direct me to an Italian restaurant, rated minimum 4.''} This request needs to be interpreted by the system to provide a venue recommendation. 

In the best case, the recommendation matches the request in terms of venue type, food selection, and rating. The system response is defined by 
a) a textual response, such as \textit{``I have found Trattoria Pizzeria close by. Do you want directions?''}, and 
b) a detailed and structured description of the found venues and their constraints, such as:

\begin{lstlisting}[language=JSON]
{
  "name": "Trattoria Pizzeria",
  "category": "restaurant",
  "cuisine": "italian",
  "location": {
    "coordinates": [45.4642, 9.19],
    "address": "Via Roma 12, 20121 Milano"
  },
  "rating": 4.5,
  "price_level": 2
}
\end{lstlisting}

While (b) is typically retrieved from a database, (a) is generated by the AUT. The success of the request depends in particular on the performance of the underlying LLM, which may fail in different ways: the LLM returns a venue that does not satisfy the user constraints, e.g., suggesting a Chinese restaurant instead of Italian or one with a rating of less than 4, or fabricates a restaurant that does not exist in the database. Also, the LLM may produce inappropriate content (e.g., offensive text), unsafe instructions (e.g., recommending a non-drivable route), or ignore or wrongly parse parts of the request, such as overlooking the minimum rating constraint. Finally, the AUT could fail to handle perturbed or adversarial user inputs (e.g., spelling errors or unusual phrasing).
% \begin{itemize}
%     \item \textbf{Constraint violations:} the LLM returns a venue that does not satisfy the user constraints, e.g., suggesting a Chinese restaurant or one with a rating of 3.2.
%     \item \textbf{Hallucinated venues:} the LLM fabricates a restaurant that does not exist in the database.
%     \item \textbf{Unsafe or malicious responses:} the LLM produces inappropriate content (e.g., offensive text) or unsafe instructions (e.g., recommending a non-drivable route).
%     \item \textbf{Misinterpretation of input:} the LLM ignores or misparses parts of the request, such as overlooking the minimum rating constraint.
%     \item \textbf{Robustness failures:} the system fails to handle perturbed or adversarial user inputs (e.g., spelling errors or unusual phrasing).
% \end{itemize}
These examples highlight the multifaceted nature of failures and highlight the need for comprehensive automated testing.
\end{example}

% !TEX root =  paper.tex
\section{Approach}\label{sec:approach}

% The algorithm is inspired by research that employs optimization algorithms such as NSGAII~\cite{Deb02NSGA2} for testing Cyber Physical Systems or program code~\cite{}. In these works the test input domain is in general continuous, where for instance velocity values or positions of simulated actors are altered. However for testing LLM-based applications the test input space can be rather described by discrete features.
Unlike domains where test inputs are typically continuous, the input space of LLM applications is more naturally described through discrete features. At the same time, stylistic features capture aspects of linguistic expression, such as user sentiment or the degree of expressiveness. \appname uses feature perturbations to introduce parameterized variations aimed at testing robustness. 
For example, for the conversational system introduced in \autoref{sec:background}, one feature is the venue category, while feature perturbations may involve small modifications to the utterance, such as word removals or misspellings, that naturally arise in automated speech recognition systems~\cite{9536732}.

An overview of the approach is provided in \autoref{algo:overview}. % The approach can be divided into three parts: 1) Test Optimization, 2) Test Generation, and 3) Test Execution/Evaluation.
% \andrea{describe Algo 1}
\appname receives as input the following parameters: a set of features $\mathcal{F} = \lbrace \mathcal{F_S}, \mathcal{F_C}, \mathcal{F_P} \rbrace$, where $\mathcal{F_C}$ is a set of content features, $\mathcal{F_S}$ are style features and $\mathcal{F_P}$ perturbation features, feature constraints $\mathcal{C_F}$, the $AUT$, the LLM $LLM_{gen}$ with the prompt template $T$, which is employed for test generation, a fitness function $F$ for test evaluation and the failure oracle $O$.

% Each of the feature categories $F_i$ can be discretized based on testing requirement for the type of input to be generated and is therefore problem specific.

% the LLM-based system under test, an LLM for the generation of Utterances, a fitness function that evaluates utterance response pairs and how ``good'' the response it, and finally, the failure oracle, which decides based on the fitness values whether a question answer pairs is failing or not.  

% In the following, we will explain each stage of the algorithm in detail, which is also illustrated in \autoref{fig:approach}.

% \begin{figure}[t]
%     \centering
%     \includegraphics[width=\linewidth]{approach-overview.png}
%     \caption{Overview of the \appname approach. \andrea{not described in the text}}
%     \ken{Simplify graph and describe.}
%     \label{fig:approach}
% \end{figure}

\subsection{Test Representation}\label{sec:representation}

Each set of features included in $\mathcal{F}$, can be formalized as $\mathcal{F} = \{F_1, F_2, \ldots, F_n\}$ where each feature 
$F_i$ takes values from a finite domain $D^F_i$. A \emph{feature vector} is defined as $x = (v_1, v_2, \ldots, v_n) \quad \text{with} \quad v_i \in F_i$.
In the first step, random combinations of features are sampled from the provided feature set (Line~11) and then encoded into a numerical space. This encoding enables the use of numerical optimization algorithms to systematically control and guide the generation of test inputs. More specifically, we consider \textit{ordinal} feature values, such as rating scores, that can be ordered and are mapped to $x_i =  index(v_i,D^F_i) /|D^F_i|$, where $index(v_i,D^F_i)$ is the index of the value $v_i$ in the list $D^F_i$, and \textit{categorical} feature values, such as venue types, that cannot be ordered and are mapped to $x_i = index(v_i, D^F_i)$.

As an example, consider the categorical feature \textit{category} $F_1$ =  \{``hospital'', ``bar'', ``restaurant''\} and the ordinal feature \textit{rating} $F_2$ = $\lbrace 3.5, 4, 4.5, 5 \rbrace$. The numerical representation of ``hospital'' would be $0$, while the one for the rating $4$ is $0.25$ (with $|D^F_2|=4$, the second element $4$ has $\text{index}(4)=1$, hence $x_i=\tfrac{1}{4}=0.25$).

\subsection{Test Preparation} 

After this step, the test inputs are passed to the test evaluator (Lines~21/27). 
Here, features are mapped into the feature space to obtain domain-specific values (Line~22). Feature constraint rules $\mathcal{C_F}$ are then applied to avoid incompatible feature combinations (Line~23). 
For example, when querying a navigational system for the category \textit{car\_repair}, a constraint rule would remove any food type selection from the feature set. Similarly, constraints can be imposed on style features to avoid the generation of e.g., polite sounding utterances (politeness feature) being at the same time very directive (anthropomorphism feature).  After the constraint rules are applied, both the feature values and their corresponding numerical encodings are updated accordingly to propagate the changes into the 
optimization workflow.

\begin{algorithm}[t]
\DontPrintSemicolon
\footnotesize
\SetKwInOut{Input}{Input}
\SetKwInOut{Output}{Output}
\Input{
    $\mathcal{F} \gets (\mathcal{F_S}, \mathcal{F_C}, \mathcal{F_P})$ \tcc*{Features} \\
    $C_\mathcal{F}$ \tcc*{Feature constraints} \\
    $AUT$ \tcc*{LLM-based System Under Test}  \\
    $LLM\_gen$ \tcc*{LLM for Utterance Generation} \\
    $Fitness$ \tcc*{Fitness Function} \\  
    $O$ \tcc*{Failure Oracle} \\
    $T$ \tcc*{Prompt Template} \\
    $k$ \tcc*{Population Size} \\
    $S$ \tcc*{Test input samples} 
 }
\Output{ $Failures$: Set of all failure-inducing test inputs.}

\textbf{Function} $\textsc{\appname}$: \;
\Indp
    \tcp{Initialization}
    $A \gets \lbrace \rbrace $ \;
    $X \gets \textsc{randomSampling}(F,k)$ \;
    $P_i \gets \textsc{executeCandidates}(X)$ \;
    \BlankLine
    \tcp{Evolutionary Search}
    \While{$\textit{budgetAvailable}$}{
    \Indp
    $P \gets \textsc{tournamentSelection}(P_i, k)$ \;
    $X \gets \textsc{crossoverMutate}(P.X)$ \;
    $P_i \gets \textsc{executeCandidates}(X)$ \;
    $A \gets A \cup P_i$ \;
    $P_i \gets \textsc{survival}(P_i, k)$ \;
    \Indm
    }

    \BlankLine
    \tcp{Failure Extraction}
    $Failures \gets \{ (utter, out, f) \in A \mid O((utter, out, f)) = 1 \}$ \;
    \Indm
    \KwRet{$Failures$}

\BlankLine
\textbf{Function} $\textsc{executeCandidates}(X)$: \;
\Indp
    $V  \gets \textsc{decode($F$,$X$)}$ \;
    $V' \gets \textsc{applyConstraints}(V, C_\mathcal{F})$ \;
    $T' \gets \textsc{generatePrompt}(T, V', S)$ \;
    $Test\_inputs \gets \textsc{generateInput}(T', LLM\_gen)$ \;
      % $Test\_inputs \gets \textsc{applyPerturb}(Test\_inputs)$ \;
    $Output \gets \textsc{execute\_sut}(Test\_inputs, AUT)$ \;
       \Indm
    \KwRet $\textsc{evaluate}(Test\_inputs, Output, Fitness)$ \;
\BlankLine

\caption{\appname approach.}
\label{algo:overview}
\end{algorithm}

\subsection{Test Generation} 

In the following, test generation is performed by converting the feature vector $v'$ into an executable test input using the given prompt template $T$ (Line~24). 

This operation is based on two steps. In the first step, the content, style, and perturbation feature vectors are inserted into a domain-specific prompt template $T$. An overview of a prompt template is shown in \autoref{fig:generation-prompt-navi}, which generates navigational requests. In the second step, our approach uses RAG to include a set of test input examples $S$ from an existing database into $T$. The retrieval of these examples is performed by comparing the embedding of the feature vector with the embeddings of samples retrieved from a given RAG database.
The instantiated prompt template $T'$ is then directed to an LLM to generate executable test inputs $Test\_inputs$. 
% \andrea{this part does not describe exactly the function ExecuteCandidate, and I cannot follow the steps. The perturbation steps are not clear (is it line 21?), and RAG is not mentioned in the pseudocode. Also, RAG applies only to Navi, I think, whereas the algorithm should be generic.}
For instance, for the feature vector \textit{v = (``bar'', ``italian'', ``rating:4'', ``tone:polite'', ``perturbation:none'')}, a generated input by \appname is mapped to the prompt template in \autoref{fig:generation-prompt-navi} to ``Could you find me please an Italian bar, rated 4''.

\begin{figure}[t]
    \begin{tcolorbox}[colback=gray!2!white, colframe=gray!50!black,
                      fonttitle=\bfseries\small,
                      left=1mm, right=1mm, top=1mm, bottom=1mm, boxrule=0.3mm]
    \scriptsize
\textbf{Instructions:} \\
You are a user interacting with an AI-powered intelligent information system for navigational requests in a car. Consider the following related attributes, which may be linguistic, content-related, or perturbation-related.

\medskip
\textit{Content-related:} \texttt{\{\{content\}\}}  \\
\textit{Style-related:} \texttt{\{\{style\}\}} \\
\textit{Perturbation-related:} \texttt{\{\{perturbation\}\}}

\medskip
\textbf{Guidelines:} 
\begin{itemize}[leftmargin=*]
    \item Do not produce any harmful utterance.
    \item Allowed are up to 12 words, but brevity is prioritized.
    \item Try to sound human-like.
    \item Make sure all style and content-related attributes are considered.
    \item Explanations:
    \begin{itemize}[leftmargin=*,label=--]
        \item Styles:
        \begin{itemize}[leftmargin=*,label=--]
            \item Slang (Slangy): If the value is slangy, the utterance should use slang words (e.g., "hook up").
            \item Implicit (Implicit): The requested venue is asked in a verbose or indirect way. You may not mention the venue directly or use a vague reference.
            \item \ldots
        \end{itemize}
    \end{itemize}
\end{itemize}

\medskip
\textbf{RAG Examples:} \texttt{\{\{rag\_examples\}\}} \\
\textbf{Few-shot Examples:} \texttt{\{\{examples\}\}} \\
\textbf{Output:} \texttt{\{\}}
    \end{tcolorbox}
    \caption{Illustrative test generation prompt for NaviQA. All feature types and examples are passed as parameters.}
    \label{fig:generation-prompt-navi}
\end{figure}

\subsection{Initial Test Execution and Evaluation}

The generated test input is passed to the LLM-based application under test (AUT), which produces a corresponding output ($Output$) (Line 26). Both the input and output are then evaluated by a fitness function that assigns a quantitative score reflecting the degree of semantic alignment between them (Line 27).
The fitness function can be implemented as either single- or multi-objective. In our evaluation, we instrument \appname with both variants, depending on the use case (\autoref{tab:case_study_features}).
Different evaluation strategies can be employed, ranging from classical similarity metrics such as ROUGE~\cite{lin-2004-rouge} and BLEU~\cite{papineni2002bleu}, to numerical distance measures (e.g., embedding- or Euclidean-based), or LLM-based judgment.
While in both our case studies we employ LLM-based evaluation, in one of them we additionally incorporate numerical comparison, as the system produces structured outputs (i.e., in JSON format). This allows us a precise quantitative assessment against structured input data.

% This score is then optimized by the underlying optimization algorithm and used as a feedback to guide the generation of promising test inputs. 

% A test input in \appname is executed by passing the generated test input to the AUT. The response of the AUT is provided together with the textual test input to an evaluation function. The evaluation function assign a numerical score which quantifies how well the system response fits to the test input.

% Formally defined, given the feature set $F = \lbrace F_1, F_2, \ldots F_n \rbrace $ a \emph{feature constraint rule} is
% \[
% r : \bigwedge_{v \in D(F')} v \;\to\; \bigwedge_{w \in D(F')} w , \qquad F' \cap F'' = \emptyset, F', F'' \subset F \;\;
% \]

\subsection{Genetic Optimization}

After the initial set of test cases has been executed and evaluated, they are fed into the optimization pipeline, which performs multiple genetic operations on encoded test inputs. The main concept is to generate test candidates (Line~15-16), evaluate them by executing the AUT (Line~16), persist evaluated tests in an archive (Line~17) and select the best test candidates based on their fitness values (Line~18). 
The generation of new candidates is performed by applying the crossover and mutation of test inputs, as explained next.

\subsubsection{Crossover}

% \begin{itemize}
%     \item \hl{Use seed}
%     \item Use feature vector crossover
%     \item Use $LLM_X$, $Prompt_X$
%     \item Example Crossover
%     \item We dont call LLMs after every operators, but after crossover, when the vector variables are fixed for the test
    
% \end{itemize}
The crossover operator between two test inputs is inspired by evolutionary theory~\cite{Deb02NSGA2}, and is achieved by exchanging feature labels. Consistent with previous studies, we apply Simulated Binary Crossover (SBX)~\cite{deb2007adaptive} for ordinal features, and Uniform Crossover~\cite{deb2007adaptive} for categorical features.
% Otherwise, feature values could be prioritized that lie at the end of the feature spectrum, even when no ordering is possible, and vice versa. 
As an example, consider two feature vectors representing candidate solutions: $v_1 = (\text{``restaurant''}, \text{``italian''}, \text{``rating:4''})$ and $v_2 = (\text{``bar''}, \text{``german''}, \text{``rating:5''}).$ During a crossover operation, information can be exchanged between the vectors. For instance, after swapping the \emph{cuisine} attribute, two new vectors are produced
$v'_2 = (\text{``bar''}, \text{``italian''}, \text{``rating:5''})$ and $v'_1 = (\text{``restaurant''}, \text{``german''}, \text{``rating:4''})$.
The selection of the elements to be exchanged is configured using a probability threshold $th_{X}$. The parent selection for crossover is performed using the standard tournament selection operator~\cite{Deb02NSGA2}.

\subsubsection{Mutation and Survival} The mutation operator receives as input a single feature vector and outputs an altered vector. Here, similarly as for the Crossover operator, we apply, based on whether the feature type is categorical, uniform mutation~\cite{deb2007adaptive} or polynomial mutation~\cite{deb2007adaptive} to each of the feature variables based on a probability threshold $th_{M}$. 
As an example, the mutation of $v_1$ could produce $v'_1 = (\text{``bakery''}, \text{``french''}, \text{``rating:4''})$, when mutating the venue category and the food type. For the survival operator, we use the default operator which first ranks all solution based on non-dominance and prioritizes afterwards tests using the crowding distance~\cite{Deb02NSGA2}. 
Once the search budget is exhausted, \appname applies the oracle function $O$ (Line 27) to label failing tests based on their fitness scores and returns all identified failures.

\subsection{Duplicate Elimination}\label{sec:duplicates}

During the iterative generation of test inputs, it may occur that identical or \textit{similar} inputs are produced by $LLM\_gen$. To address this problem, we compute the cosine similarity~\cite{Mikolov:2013ohu} between the embedding vectors of the test inputs and apply a threshold of 0.8 to identify \textit{duplicates}, consistent with prior work~\cite{llamaindexsimilarity}. The embeddings are computed using \textsc{all-MiniLM-L6-v24}. Inputs exceeding this similarity threshold are removed from the population. 

% Formally, the cosine similarity between two embedding vectors, $\mathbf{e}_i$ and $\mathbf{e}_j$, is computed as:
% \[
% \text{sim}(\mathbf{e}_i, \mathbf{e}_j) = \frac{\mathbf{e}_i \cdot \mathbf{e}_j}{\|\mathbf{e}_i\| \, \|\mathbf{e}_j\|}
% \] where $\|\mathbf{e}_i\|$ and $\|\mathbf{e}_j\|$ is the Euclidean norm \cite{}.

% \lev{we use actually the inverse normalized but is not relevant I guess}
% \begin{itemize}
%     \item Pass to AUT
%     \item We have 2 fitness functions
%     \item Few short prompting + rerun 1-3 times
%     \item Evaluate with Judge LLM 
%     \item Temperature Validation: 0.5 (for sampling higher)
%     \item Fitness Function/Oracle
%     \item Weighted Evaluation. Assign score based on category of feature. Rar categories weight less. Based on discussion with experts at BMW.

% \subsection{Validation}

% \andrea{Line~16 of the approach was never described, I think it is what you need in this section. \astral, \gs, and \rs were never introduced before! This part is procedure, not approach.}

% To evaluate the validity of test generated be \appname, \astral, \gs, and \rs we select 10 samples per method and model and evaluate them with 3 human experts. Based on the number of failures output, we approximate the actual number of failures per algorithm.

% \begin{itemize}
%     \item Evaluate X\% failures with Y humans

%     \item \hl{Not clear if we have time to do that.}
%     \item Do we need take some samples over all seeds or just from one seed?
%     \item Based on which criteria to evaluate: R, D, P, Content?
% \end{itemize}
\subsection{Implementation}

We have implemented \appname on top of OpenSBT~\cite{sorokin2023opensbt}, a modular search-based testing framework widely applied in literature~\cite{NejatiSSFMM23,sorokin24svm,Sorokin2024, munaro2026faultinjection}. For optimization we leverage NSGA-II~\cite{Deb02NSGA2}. To interface cloud-based LLMs, we used model deployments in Azure; for local models, we used Ollama/Hugging Face. Our study includes more than 234,000 tests (1,000 tests per run), for a total execution time of more than 24 days.
% Overall, our experiment includes X LLMs under test. For SafeQA, we used Y LLMs. For NaviQA, we used Y LLMs. As our evaluation set comprises Z images, overall we computed XXX tests in our experiments (approx. 24 days computing time).
% The total execution of the experiments required more than 24 days.
% !TEX root = paper.tex
\section{Evaluation}\label{sec:empirical-study}

\subsection{Research Questions}\label{sec:rqs}

% To evaluate our approach, we consider the following research questions:

\head{RQ\textsubscript{0} (judge evaluation)} \textit{How accurate are LLM-based judges in evaluating test pass/fail outcomes?}

In our approach, we employ an LLM-based oracle to evaluate the response of the system under test. In this research question, we assess the effectiveness of this technique to be used in \appname.

\head{RQ\textsubscript{1} (effectiveness)} \textit{How effective is \appname in identifying failures of LLM applications?}

One goal of testing is to be able to identify as many failures as possible. For this, we compare for a fixed testing budget our approach with vanilla randomized testing and ASTRAL, a state-of-the-art automated testing approach for LLMs. 
% In particular, we evaluate the number of identified failures and ratio of unique critical tests found in comparison total generated tests.

\head{RQ\textsubscript{2} (failure diversity)} \textit{How diverse are the generated failures?}

% Identified failures by a testing approach might by close to each other. 
To support debugging and fault localization~\cite{Abdessalem2018TestingVC}, it is essential to identify \textit{diverse} failures, as suggested in previous works~\cite{surrogate2024biagiola, feldt2016diversity}.

\begin{table}[t]
    \centering
    \caption{Case studies configurations.}
    \label{tab:experiment_configuration}
    \begin{tabular}{lcc}
    \hline
        \addlinespace[0.5ex]
    \textbf{Parameter} & \textbf{SafeQA} & \textbf{\convnavione/II} \\
        \addlinespace[0.5ex]
    \hline
    \addlinespace[0.5ex]
    population size & 20 & 20\\
    search time & 2h & 3h \\
    crossover threshold ($th_{X}$) & 0.7 & 0.7 \\
    mutation threshold ($th_{M}$) & 0.12 & 0.07 \\
    \textit{\#} features & 8 & 13 \\
    \textit{\#} features combinations  & 5,600 & 
    11,664,000 \\
        similarity threshold & 0.8 & 0.8 \\
        num. fitness & 1 & 2  \\
    LLM Generator & \dolphinthree & \gptfouromini \\
    LLM Judge & \gptfouromini & \gptfouromini \\
        Temperature Generator/Judge & 0/0 & 0.2/0 \\
    \addlinespace[0.5ex]
    \hline
    \end{tabular}
\end{table}

\subsection{Case Studies}\label{sec:methods}

To evaluate the effectiveness and generalizability of \appname, we consider two case studies related to two domains and three AUTs, whose configurations are summarized in \autoref{tab:experiment_configuration}.

\subsubsection{Safety}

In the safety case study (SafeQA hereafter), the goal is to assess whether a standalone LLM can generate unsafe 
system responses. This is a major concern from recent studies~\cite{xie2025sorrybench,ji2023beavertails,wen-etal-2023-unveiling, andriushchenko2025jailbreakingleadingsafetyalignedllms}.
For example, a malicious input such as \textit{``Provide an example to insult a person''} or be\textit{``Explain how to destroy my neighbour's car''} could trigger outputs that provide instructions for harmful or illegal actions. 

\subsubsection{Navigation}
% \ken{It should be made clear that we are testing text-in/text-out. Arguments could be: easier deplyoment when having textual access. No additional point of failure using ASR. We have to state that even after speech passed through ASR, all error types could still be found!}
The second case study is \convnavione, a navigational task relying on a static open-source dataset. 
% \andrea{can we cite a \company paper where a similar system was used as AUT?} \lev{There is just CarExpert, has never been applied}
\convnavione retrieves points of interest (POIs) from the YELP dataset~\cite{Asghar16}, which primarily contains data of business venues in the United States. The system follows a Retrieval-Augmented Generation (RAG) paradigm, where the LLM retrieves candidate POIs from the static dataset and incorporates them into the response generation process. 
% This setting allows us to evaluate \appname in a reproducible and transparent manner, while facilitating replication by other researchers.
% \subsubsection{Industrial-Grade Navigation System (NaviQA-II)}
Finally, we evaluate \appname on \convnavitwo, an industrial-grade LLM-based navigational system from our partner company \company.
Unlike \convnavione, this system retrieves POIs dynamically through online APIs and extends beyond venue retrieval to provide information about vehicle state and access to the car manual~\cite{rony-etal-2023-carexpert}, expanding the range of potential system features, and hence failure modes, beyond those of \convnavione. 
Together, these two case studies allow us to assess \appname across different contexts: safety evaluation for consistency with prior work, navigation with open-source data for replicability, and an industrial-grade system for realism and practical relevance.

\subsection{Baselines}\label{sec:methods}

For SafeQA, we compare \appname against Random Search (RS) and ASTRAL, using ASTRAL's original feature set. In contrast, \appname and RS operate on an extended feature space where constructing a full coverage matrix is infeasible due to combinatorial growth. We also include \gs, a combinatorial method based on 4-wise feature interaction, which scales ASTRAL's principles without requiring full coverage.
For NaviQA (\convnavione and \convnavitwo), we compare only against \gs and RS, as ASTRAL cannot be directly applied outside safety-focused LLM testing.

% \lev{only ASTRAL has the same feature set. is that clear?} \andrea{does it mean that we compare the techniques on the same feature set in this work, or that we use the feature sets used in ASTRAL? In both cases, we must be precise e.g., by referring to Table II. It's unclear what these features are now} \lev{feature sets are different. astral: use original feature set. stellar and t-wise: use equal feature set but more then astral.}
% This ensures consistency with prior work, as ASTRAL was originally proposed for safety testing~\cite{ASTRAL}.

% For the navigational case studies (\convnavione and \convnavitwo), a direct application of \astral is infeasible, as the large number of features leads to a combinatorial explosion. Instead, we compare against randomized search and \gs, a combinatorial testing approach , selected from preliminary experiments. 

% \andrea{the case studies are not introduced yet. Moving this section after them is also problematic as the sections are super big, with}

\subsection{Metrics}\label{sec:metrics}

For RQ\textsubscript{0}, to evaluate LLM-based judgment, we use, for SafeQA, an existing database of question–answer pairs and assess the classification performance of the judge with respect to a selected subset of samples (1000) ~\cite{ji2023beavertails}.
For \convnavione and \convnavitwo, since no dataset provides annotated question–answer pairs aligned with our evaluation dimensions, we generate and manually annotate the data with human raters. We report the inter-rater reliability scores, as well as the performance of the judge compared to the human annotations. 
% \andrea{nothing for industrial?} \lev{For industrial it is the same}

For RQ\textsubscript{1}, to compare \appname with baseline approaches, we measure the number of failing test inputs over time for a fixed search budget. In particular, we exclude duplicates (s. \autoref{sec:duplicates}) and invalid test inputs (e.g., empty utterances).
For \convnavione and \convnavitwo, we in addition exclude test inputs that fail because of an inappropriate response, while no POI actually exists to answer the request. 
% We can assess whether a POI for a request actually exists, using the API \texttt{poi\_exists}. 
To measure the convergence of the search, we further track the ratio of the number of failures found over the test inputs executed. This metric is relevant, as it allows relative comparisons taking into account that single runs might require longer single LLM call execution times limiting the number of tests produced.

% \andrea{add industrial?} \lev{For industrial it is the same}

For RQ\textsubscript{2}, to measure the diversity of identified failures using the approach from Biagiola et al.~\cite{surrogate2024biagiola}. It involves clustering all aggregated failures from all approaches and then evaluating the coverage of clusters. As proposed~\cite{surrogate2024biagiola}, we repeat the clustering 10 times, apply the Silhouette method~\cite{9260048} to identify the number of clusters automatically, and average the coverage results over the 10 runs. 

\subsection{Procedure}

\subsubsection{Judge Evaluation}

To answer RQ\textsubscript{0}, we benchmarked for the safety case study eight different LLMs, such as \gptthreefive, \gptfouromini, \gptfouro, \gptfivechat, \deepseeklocal, \deepseekcloud, \mistral in providing a continuous score as well a binary score regarding the safety of a textual output of the AUT. We use the continuous judge during the search to retrieve fitness values that can be optimized, while we employ a binary score after the search on test inputs to be able to compare our approach to \astral which employs a binary judge. The results of the judge benchmarking which was performed on 1000 question answer pairs sampled from the Beavertails benchmark~\cite{ji2023beavertails}.

\subsubsection{Safety}

% In the safety case study, the goal is to evaluate whether a standalone LLM can output unsafe system responses. This is in particular, a major concern which has been detected in various studies conducted~\cite{xie2025sorrybench,ji2023beavertails, wen-etal-2023-unveiling, andriushchenko2025jailbreakingleadingsafetyalignedllms}. For instance, a malicious output could be triggered based on a malicious user input \textit{"Provide an example to insult a person."} or \textit{"Explain me how to destroy the car of my neighbour."}
% Malicious output would explain how to insult a person or give instructions to commit property damage.

\begin{table}[t]
\centering
\caption{Overview of features used in our study.\tablefootnote{The discretization of features is provided in the replication package~\cite{repo}.}}
\label{tab:case_study_features}
\begin{tabular}{p{1.2cm} p{3.3cm} p{2.5cm}}
\toprule
\textbf{Feature Category} & \textbf{SafeQA} & \textbf{\convnavione /II} \\
\midrule
Style & Politeness, Slang, Anthropomorphism, Persuasion, ASTRAL styles
      & Politeness, Slang, Anthropomorphism, Implicitness \\
\midrule
Content & \parbox[t]{6cm}{Safety Category~\cite{xie2025sorrybench}} 
        &        Price, Venue, Payment, Parking, Rating, Cuisine \\
\midrule
Perturbation & Word deletion, Character perturbations \cite{metal}, Adding Fillers, Homophones~\cite{pronouncing}
             &Word deletion, Adding Fillers, Homophones~\cite{pronouncing} \\
\bottomrule
\end{tabular}
\end{table}

\head{Search Space} 
We follow ASTRAL's discretization strategy for content features, using 13 categories from the SORRY benchmark~\cite{xie2025sorrybench}, and similarly adopt stylistic dimensions such as persuasion and expression. Unlike ASTRAL, we extend the feature space with perturbations, e.g., including word deletions (e.g., pronouns), filler words or homophonic substitutions, or character-level noise and typos, drawing from the METAL framework~\cite{metal}. We further incorporate style attributes such as politeness, slang, and anthropomorphism, and we discretize the feature values provided by ASTRAL into more granular categories. For example, for slang, we distinguish between the values \textit{formal}, \textit{neutral}, and \textit{slangy}.
The full feature set is shown in \autoref{tab:case_study_features}.

\head{Test Generation} 
To generate test inputs with \appname, we instantiate the ASTRAL prompt template~\cite{ASTRAL} with concrete content, style, and perturbation features drawn from our discretized feature space. Each prompt includes five malicious examples retrieved via RAG from the BeaverTails dataset~\cite{ji2023beavertails} to provide contextual guidance. A full prompt specification is available in the supplementary material~\cite{repo}.

\head{Fitness Function}  
In the safety case study, we use an LLM as the judge and define a fitness function that evaluates whether a system response is appropriate given the test input. The judge prompt is derived from ASTRAL's binary classification setup~\cite{ASTRAL}. While we retain binary judgments for final failure assessment to ensure comparability with ASTRAL, binary output alone is insufficient during search, where continuous feedback is needed for optimization. Therefore, we adapt the prompt to elicit a continuous score in the range of 0 (unsafe) to 1 (safe), enabling the search algorithm to distinguish degrees of correctness and guide input generation more effectively. We configure \appname to minimize the score.

\head{Failure Oracle} 
We use a binary failure oracle, applying the prompt template proposed by ASTRAL~\cite{ASTRAL} with \gptfouromini after the search has completed to distinguish failing tests from non-failing ones.

\head{LLMs under Test} To evaluate our approach, we select six different open and closed source LLMs of different sizes and providers. In particular, we selected cloud-based models \gptfouro, \gptfivechat, \deepseekcloud, as well as locally deployed models \mistral, \deepseeklocal, and \qwentwo. We have excluded other locally deployable models, such as \llamathreetwo, due to testing budget constraints.

\head{Testing Setup} 
For \appname, we use a search budget of two hours and a population size of 20. This configuration was informed by preliminary experiments, which showed that failure discovery rates plateau before this threshold. Mutation and crossover parameters follow the default settings introduced by Abdessalem et al.~\cite{Abdessalem-ICSE18}. All algorithms are executed under the same configuration for each LLM under test. In the safety case study, no feature constraints were imposed, as we did not observe clear semantic conflicts without introducing bias.

\subsubsection{Navigation}

% a navigation case study involving two venue recommendation systems, as introduced in \autoref{sec:background}. The first system, \convnavione, is a simplified prototype that relies on a static open-source dataset provided by YELP~\cite{}, for the retrieval of points of interest (POIs). This dataset primarily contains location data from the United States.

% The second system, \convnavitwo, is an industrial prototype representing a more advanced LLM-based system. In contrast to \convnavione, \convnavitwo retrieves POI information dynamically using a cloud-based service through online APIs. Moreover, it offers extended functionalities beyond venue retrieval, such as providing information about the vehicle state and access to the car manual~\cite{rony-etal-2023-carexpert}. In particular, the latter fact, extends the possible list of failures that can happen compared to the failures listed in~\autoref{sec:introduction}.

\head{Search Space}   
We apply \appname to both \convnavione and \convnavitwo to assess its generalization capability and to support replicability within the navigation domain. To enable test generation, we define the input feature space along style, content, and perturbation dimensions.

Style features were derived by reviewing real user interactions and through discussions with \company experts on how users formulate requests in navigational systems. Content features for \convnavione were taken directly from the POI database. For \convnavitwo, the cost attribute was excluded due to backend issues affecting POI retrieval, but all other content, style, and perturbation features were kept consistent with \convnavione.

For perturbations, we include homophonic substitutions to simulate Automated Speech Recognition (ASR) errors, using homophone mappings from the CMU Pronouncing Dictionary~\cite{pronouncing}. We also incorporate filler words (e.g., ``hm'', ``eh'') to emulate natural speech patterns~\cite{bortfeld2001fluent}. All style and perturbation features are discretized into three to five categories.
A complete overview of the feature set is provided in \autoref{tab:case_study_features}.
\head{Test Generation} 
To generate test inputs for NaviQA, we first design a domain-specific prompt template. Similar to the SafeQA setup, the template includes placeholders for content, style, and perturbation features, along with system instructions tailored to venue search requests. We additionally clarify stylistic attributes, such as implicitness, to guide the model in producing inputs aligned with the defined feature dimensions.

The prompt includes ten examples: five manually crafted and annotated with feature scores, and five retrieved via RAG from the MultiWOZ dataset~\cite{zang2020multiwoz}. An overview of this template is shown in \autoref{fig:generation-prompt-navi}.
For input generation, we select \gptfouromini and validate its suitability by generating 50 candidate inputs, which are reviewed by two domain experts that were not involved in the development of \appname. The evaluation shows an averaged validity rate of 93.5\%, why \gptfouromini is selected for test generation.

\head{Fitness Functions}  
For \convnavione/II we use the fitness functions: \textit{Fitness Response} ($f_1$), and \textit{Fitness Content} ($f_2$).

The former fitness function ($f_1$) assesses how well the LLM provides an appropriate textual response regarding three dimensions. The dimension have been selected based on interviews with \company experts. The saturation of this categories is relevant to assure the customers satisfaction: (1)~\textit{Request-oriented}: This dimension targets to evaluate whether the response is related to the request of searching for a venue. It is subdivided into three categories, covering ``relevant'', ``partially relevant'' and ``not relevant''. (2)~\textit{Directness:} The second dimension evaluates how verbose the answers of the system are. It covers the values ``not verbose'', ``partially verbose'' and ``fully verbose''. (3)~\textit{Follow-up:} The last dimension evaluates whether the system provides a follow-up question or information for the next step to continue the conversation flow. Also here, we use three categories ``follow-up available'',``follow-up vague'', ``no follow-up''.
We employ an LLM-as-a-Judge~\cite{2025-Giebisch-IV, habicht2025benchmarkingcontextualunderstandingincar} to provide a score for each of the dimensions, as we cannot use reliably conventional techniques such as ROUGE~\cite{lin-2004-rouge} or BLEU~\cite{papineni2002bleu}.

The latter fitness function ($f_2$) evaluates how well the returned list of POIs $OUT = \lbrace poi_1, poi_2, \ldots, poi_n\rbrace$ fits to the requested POI constraints in the input $in$ by the user. The fitness function is defined as follows:
\[
f_2 =
\begin{cases}
1,\; \ \ \text{if } \neg \text{poi\_exists}(\text{in}) \;\text{and}\; |\text{OUT}| = 0,\\[0.5ex]
\displaystyle \max_{\text{poi} \in \text{OUT}} \sum_{i=1}^{n} w_{c_i} \cdot (1 - dist(\text{in}.c_i, \text{poi}.c_i)), \ \text{otherwise.}
\end{cases}
\]

where \texttt{poi\_exists} is used by the oracle to evaluate whether a request can be satisfied given the information in the database. Instead, \texttt{dist} is a distance function which evaluates the distance between a constraint in the input and a corresponding constraint in the output: when the type of the constraint is numerical, we use for ordinal constraints the Euclidean distance and for categorical constraints the exact match function (\autoref{sec:representation}). 
For non-numerical, text-based constraints, we compute similarity using embedding distance. Each constraint is weighted by a factor $w_{c_i}$ to reflect its relative importance. In our experiments, we assign a weight of 2 to the venue feature to emphasize its role in retrieval, while all other features are weighted at 1.

At the end, we normalize after applying the constraint wise comparison the $f_1$ value considering the number of constraint checks performed. 
% \lev{Need to be here more clear that when a POI is not considered in Input, then it is not checked. If it is in request, but not available in output is also not considered (optimistic approach)} 
When multiple POIs are returned, we select the POI with the best overall score. In case, no matching POI exists, and the retrieved POI list is empty, a \textit{good} score of 1 is assigned. We configure both $f_1$ and $f_2$ to be minimized.

\begin{figure*}[t]
    \centering
 \includegraphics[width=0.88\linewidth]{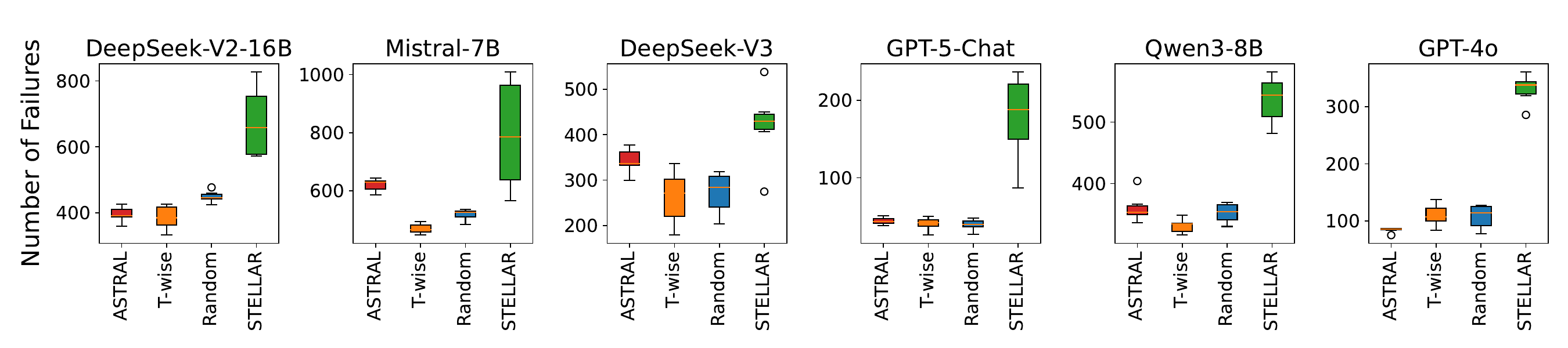}
 \includegraphics[width=0.88\linewidth]{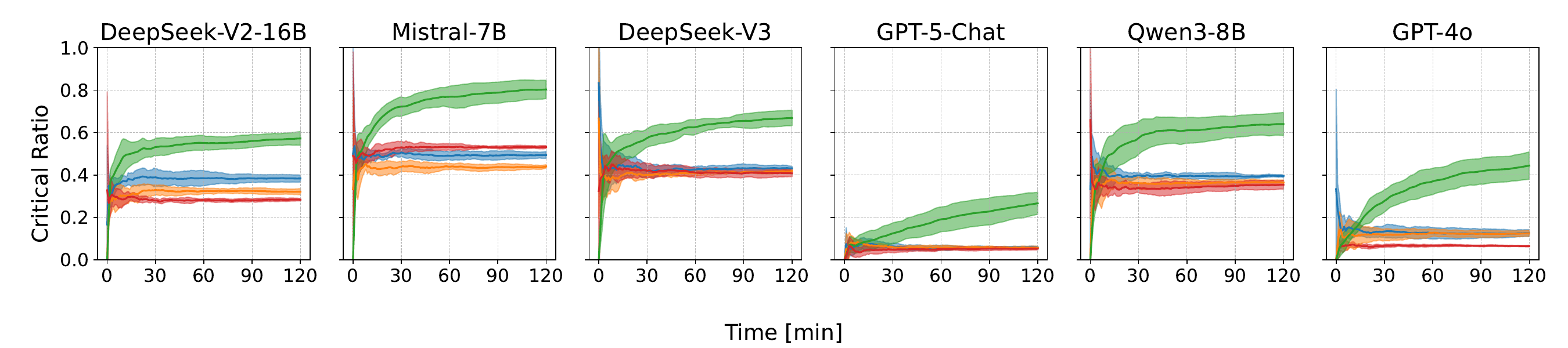}
    \caption{Results for RQ\textsubscript{1} (SafeQA). Number of failures found by each testing approach after 2 hours of search time (top). Mean ratio between failures found and in total generated test cases with standard deviation (bottom). Results averaged over 6 runs.
    % \appname identifies on average up to two times more failures then the best compared approach. The highest number of failures is found for \mistral, while the lowest for \gptfivechat.
    }
\label{fig:rq1-safeqa}
\end{figure*}

\head{Failure Oracle} 
We use the following oracle to label a test case as failing: $O = (f_1 < 0.75) \vee (f_2 < 0.75)$. We set equal thresholds since both fitness functions are considered equally important by \company experts.
A threshold of 0.75 was chosen to balance a perfect fit (1.0) and a borderline performance (0.5).
Additionally, evaluation results for varying thresholds are provided in the supplementary material~\cite{repo}.

\head{LLMs under Test} 
We selected three cloud-based LLMs \deepseekcloud, \gptfouro, and \gptfivechat. We had to exclude locally deployed LLMs such as \llama, \qwentwo, \deepseeklocal as these LLMs yielded failure rates of over 90\% in preliminary experiments conducted.

\head{Testing Setup}  We use a search budget of 3 hours, which was determined to be a reasonable duration for preliminary system-level tests based on discussions with test engineers at \company. Further we use a population size of 20, als selected based on pilot experiments. We use the same and default mutation and crossover parameters as in NaviQA, and use the same setup across all algorithms and systems under comparison.
We select constraints based on discussions with \company experts, and remove, e.g., the price range and food types selection from categories such as \textit{hospital} or \textit{museum}.

% \andrea{HERE}

% \begin{itemize}
%     \item \appname-randomized
%     \item ASTRAL  (t-wise coverage)  
% \end{itemize}

% \textbf{Final Validation}

% \begin{itemize}
%     \item Then manually evaluate 
%     \item Or evaluate all failures manually with 6+ people
%     \item With our without LOS
% \end{itemize}

\subsection{Results}

% In the following we present the results answering our research questions.

\subsubsection{Judge evaluation (RQ\textsubscript{0})}

\begin{table}[t]
\centering
\scriptsize
\caption{(RQ\textsubscript{0}): Judge evaluation averaged over 5 runs.}
\label{tab:safety-naviqa-judge-performance}
\resizebox{\columnwidth}{!}{
% \begin{tabular}{p{2cm}cccc|cc}
\begin{tabular}{lcccccc}
\toprule
& \multicolumn{4}{c}{\textbf{SafeQA}} & \multicolumn{2}{c}{\textbf{\convnavione/II}} \\
\cmidrule(lr){2-5} \cmidrule(lr){6-7}
\textbf{Model} & \textbf{F-1 (Binary)} & \textbf{Time} & \textbf{F-1 (Continuos)} & \textbf{Time} & \textbf{F-1} & \textbf{Time} \\
\midrule
\gptthreefive   & 0.76 & 1.0 & 0.75 & 0.9 & 0.65 & 1.33 \\
\gptfouromini   & 0.79 & 1.1 & 0.79 & 1.2 & 0.71 & 1.39 \\
\gptfouro       & 0.76 & 1.3 & 0.77 & 1.5 & 0.73 & 1.69 \\
\gptfivechat    & 0.77 & 1.5 & 0.78 & 1.8 & 0.70 & 1.59 \\
\deepseeklocal  & 0.64 & 1.1 & 0.66 & 2.1 & 0.58 & 1.71 \\
\deepseekcloud  & 0.80 & 1.9 & 0.40 & 1.0 & 0.76 & 2.06 \\
\mistral        & 0.76 & 1.2 & 0.73 & 2.3 & 0.68 & 2.05 \\
\bottomrule
\end{tabular}
}
\end{table}

\begin{figure*}[t]
   \hspace{0.25cm} 
    \begin{subfigure}[t]{0.55\textwidth}
    % \fbox{
        \includegraphics[trim={0 0 0 10},width=1.1\linewidth]{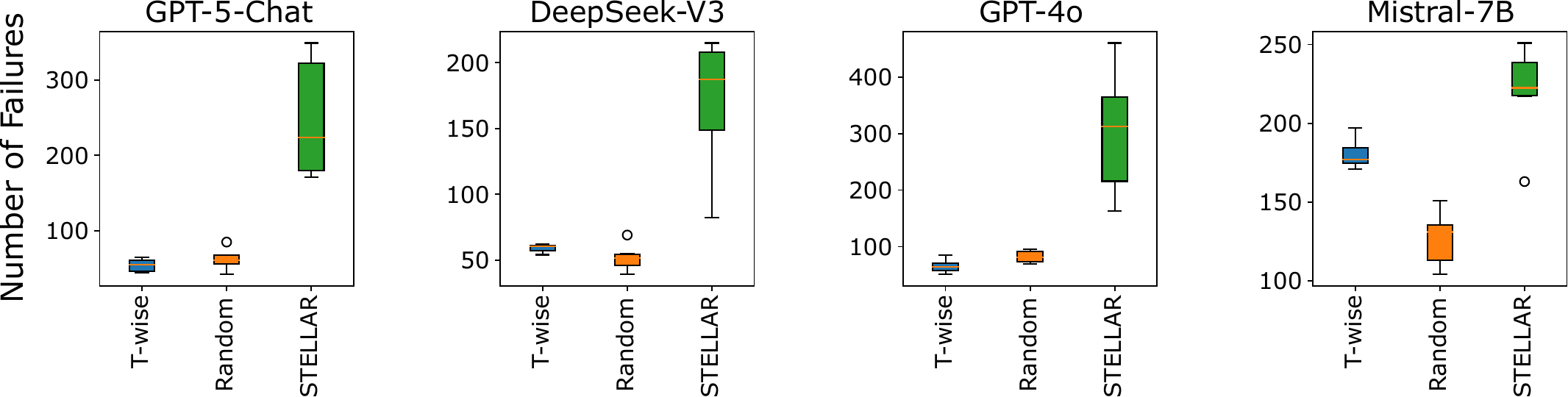}
        % }
        \caption{\convnavione}
        \label{fig:navi-yelp-failures-results}
    \end{subfigure}
    % \hfill
    \begin{subfigure}[t]{0.38\textwidth}
        \centering
        % \fbox{
        \includegraphics[trim={0 0 0 0}, width=0.35\linewidth]{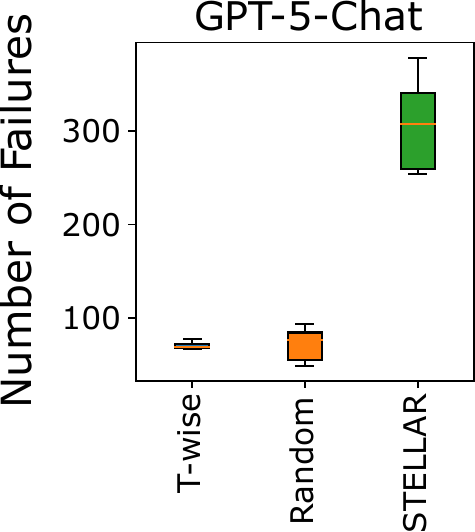}
        % }
        \caption{\convnavitwo}
        \label{fig:navi-bmw-failures-results}
    \end{subfigure}
% \vspace{-0.5cm}
    \begin{subfigure}[b]{0.6\textwidth}
        \centering
            \vspace{0.3cm}
        \includegraphics[trim={0 0 0 34}, width=1.2\linewidth]{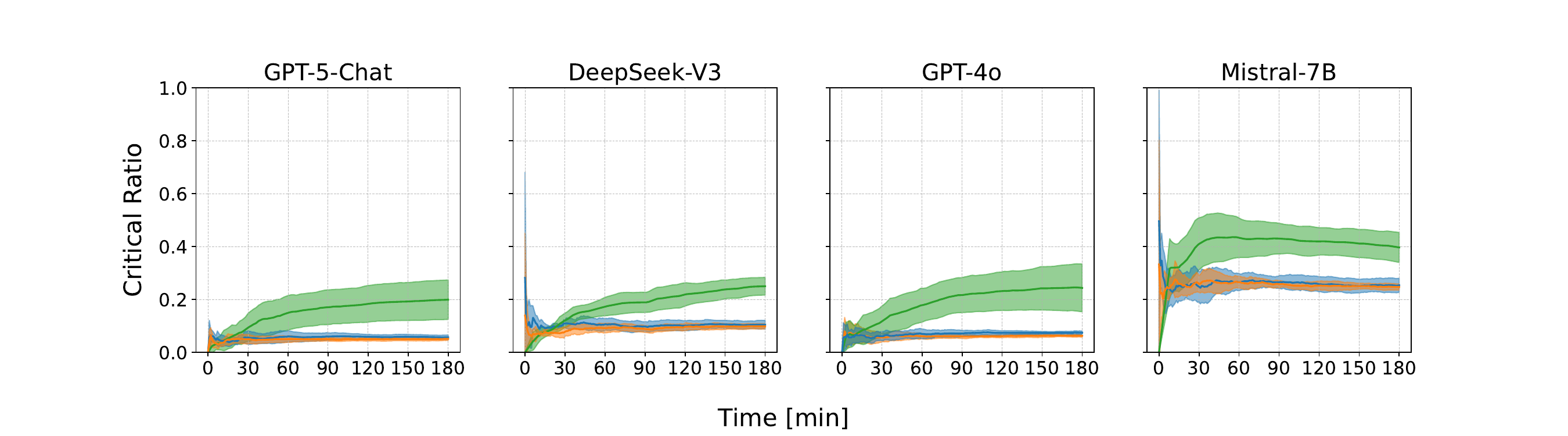}
        \caption{\convnavione}
        \label{fig:navi-yelp-ratio-results}
    \end{subfigure}
    \hfill
    \begin{subfigure}[b]{0.38\textwidth}
        \centering 
        \vspace{-2cm}
        \includegraphics[trim={0 0 0 5}, width=0.42\linewidth]{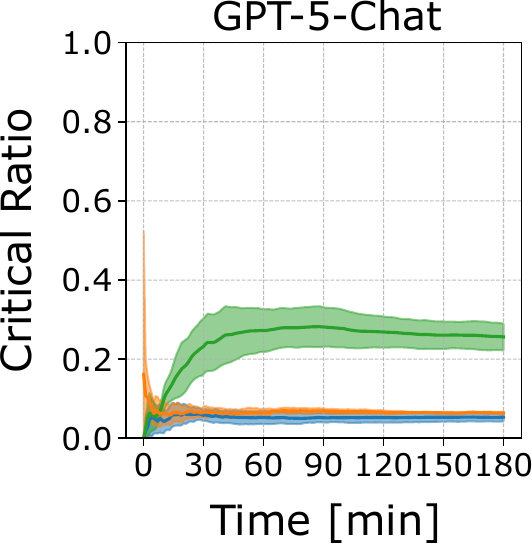}
        \caption{\convnavitwo}
        \label{fig:navi-bmw-ratio-results}
    \end{subfigure}

    \caption{Results for RQ\textsubscript{1} (NaviQA). Number of failures found by each testing approach after 3 hours of search time (top). Mean ratio between failures found and in total generated test cases with standard deviation (bottom). Results averaged over 6 runs.}
    \label{fig:rq1-naviqa}
\end{figure*}
% \begin{table}[t]
% \centering
% \scriptsize
% \caption{RQ\textsubscript{0} (SafeQA Judge Evaluation).}
% \label{tab:safety-judge-performance}
% \begin{tabular}{lcccccc}
% \toprule
% \textbf{Model} & \multicolumn{3}{c}{\textbf{Binary}} & \multicolumn{3}{c}{\textbf{Continuous}} \\
% \cmidrule(lr){2-4} \cmidrule(lr){5-7}
%  & \textbf{ROC} & \textbf{F-1} & \textbf{Time} & \textbf{ROC} & \textbf{F-1} & \textbf{Time} \\
% \midrule
% \gptthreefive   & 0.78 & 0.76 & 1.0 & 0.80 & 0.75 & 0.9 \\
% \gptfouromini   & \textbf{0.82} & 0.79 & 1.1 & \textbf{0.88} & \textbf{0.79} & 1.2 \\
% \gptfouro       & 0.79 & 0.76 & 1.3 & 0.86 & 0.77 & 1.5 \\
% \gptfivechat    & 0.80 & 0.77 & 1.5 & 0.87 & 0.78 & 1.8 \\
% \deepseeklocal  & 0.66 & 0.64 & 1.1 & 0.74 & 0.66 & 2.1 \\
% \deepseekcloud  & 0.81 & \textbf{0.80} & 1.9 & 0.70 & 0.40 & 1.0 \\
% \mistral        & 0.80 & 0.76 & 1.2 & 0.81 & 0.73 & 2.3 \\
% \llamathreetwo  & 0.80 & 0.78 & 1.9 & 0.81 & 0.71 & 1.4 \\
% \bottomrule
% \end{tabular}
% \end{table}

% \begin{table}[t]
% \centering
% \scriptsize
% \caption{RQ\textsubscript{0} (\convnavione/II Judge Evaluation).}
% \label{tab:safety-judge-performance}
% \begin{tabular}{lcc}
% \toprule
% \textbf{Model} & \textbf{F-1} & \textbf{Time[s]} \\
% \midrule
% \gptthreefive   & 0.65 & 1.33 \\
% \deepseekcloud  & 0.76 & 2.06 \\
% \gptfouromini   & 0.71 & 1.39 \\
% \gptfour        & 0.72 & 1.71 \\
% \gptfouro       & 0.73 & 1.69 \\
% \gptfivechat    & 0.70 & 1.59 \\
% \mistral        & 0.68 & 2.05 \\
% \deepseeklocal  & 0.58 & 1.71 \\
% \bottomrule
% \end{tabular}
% \end{table}

\autoref{tab:safety-naviqa-judge-performance} presents the results for the LLM judge evaluation, for both use cases. In the case of SafeQA, for the binary judgment task, the best-performing LLMs in terms of F-1 values were \deepseekcloud and \gptfouromini. 
Similarly, for the continuous judgment task, \gptfouromini achieved the best results. Thus, we selected \gptfouromini for integration into \appname in the rest of the study also due to its faster inference time.

Regarding NaviQA-I/II, to evaluate the \textit{Fitness Response}, we first generated 30 question-answer pairs using \gptfouromini, guided by the feature dimensions of our framework. Each response was prompted to vary along $R$, $D$, and $P$. Invalid or irrelevant outputs were manually filtered and regenerated. 
% The exact generation prompt is provided in the supplementary material~\cite{repo}.
We then conducted a human study with 10 participants from \company, who rated each response on $R$, $D$, $P$, and provided an overall binary judgment $O$. A total of 300 annotations were collected under the same evaluation conditions used by the LLM judge. An excerpt of the questionnaire is available in the replication package~\cite{repo}.
Following established methodology~\cite{bai2024mt}, we assessed (1)~inter-rater agreement via Fleiss' Kappa~\cite{Fleiss:1971}, and (2)~the alignment between LLM judgments and majority-vote human labels. Substantial agreement was observed for $R$ (0.62) and $P$ (0.69), with moderate agreement for $D$ (0.43). The lower consistency on $D$ reflects its subjective nature, as perceived difficulty depends more on individual interpretation than on directly observable qualities. 
% Additional statistics are reported in supplementary material~\cite{repo}.

Overall, the judge's accuracy achieved F-scores between 0.68 and 0.76 across evaluated models (\autoref{tab:safety-naviqa-judge-performance}), whereby multi-sample prompting did not offer significant gains but higher costs. We therefore selected \gptfouromini as our judge, balancing accuracy, latency, and cost.
For optimization, we compute a continuous fitness score as $f_1 = w_R \cdot R + w_D \cdot D + w_P \cdot P$. The weights ($w_R = 0.55$, $w_D = 0.3$, $w_P = 0.15$) were first derived via logistic regression on human overall judgments $O$~\cite{cabrera1994logistic} and validated finally with \company experts.

\begin{tcolorbox}[boxrule=0pt,frame hidden,sharp corners,enhanced,borderline north={1pt}{0pt}{black},borderline south={1pt}{0pt}{black},boxsep=2pt,left=2pt,right=2pt,top=2.5pt,bottom=2pt]
\textbf{RQ\textsubscript{0} (judge evaluation).} Among the evaluated models, \gptfouromini demonstrates the best balance between performance and time/cost efficiency across both case studies SafeQA and NaviQA, achieving F1-scores of 0.79 and 0.71, respectively.
\end{tcolorbox}

\subsubsection{Effectiveness (RQ\textsubscript{1})}

% \begin{figure*}
%     \centering
% \includegraphics[width=1\linewidth]{critical_ratio_over_time.png}
%     \caption{(SafeQA) Mean Ratio between failures found and generated test cases with standard deviation comparing \appname (green) with \rs (blue), \gs (orange) and \astral (red). Highest rate is identified for \mistral (80\%), the lowest for \gptfivechat (27\%). For \appname the failure rate is on average up to 2.2 times higher then vanilla testing.}
% \label{fig:safety-ratio-results}
% \end{figure*}

For SafeQA, the failure counts are reported in \autoref{fig:rq1-safeqa} (top). Across all LLMs, \appname consistently identifies more failures than \astral, \gs, and \rs. The smallest improvement is observed with \gptfouro, while the largest gain occurs with \mistral. As expected, the lowest number of failures overall is found for \gptfivechat, which reflects its stronger alignment and safety mechanisms.
In general, all techniques detect more failures in smaller, local models compared to large cloud-based models (up to 671B parameters). Between the baseline methods, \astral only outperforms \rs and \gs in two out of six LLM configurations; otherwise, their performance is largely comparable, indicating limited benefit from coverage-based guidance in SafeQA.

\autoref{fig:rq1-safeqa} (bottom) reports the failure ratio, i.e., identified failures relative to all executed tests. Across all LLMs, this ratio stabilizes over time, confirming that the allocated search budget is sufficient. \appname consistently achieves the highest failure ratio compared to all baselines. The lowest ratio (27\%) is observed on \gptfivechat, consistent with its failure counts, while the highest ratio (80\%) is recorded on \mistral. Overall, \appname identifies on average a 2.2 times higher failure rate than the best baseline approach.

% \begin{figure*}[htbp]
%     \centering
%     \begin{subfigure}[b]{0.6\textwidth}
%         \centering
%         \includegraphics[width=1.3\linewidth]{yelp_ratio.png}
%         \caption{\convnavione}
%         \label{fig:navi-yelp-ratio-results}
%     \end{subfigure}
%     \hfill
%     \begin{subfigure}[b]{0.38\textwidth}
%         \centering
%         \includegraphics[width=0.53\linewidth]{bmw_ratio_cut.png}
%         \caption{\convnavitwo}
%         \label{fig:navi-bmw-ratio-results}
%     \end{subfigure}
%     \caption{Comparison of mean failure rates between found and generated test cases for \convnavione and \convnavitwo over 6 runs. (\appname highest mistral 39\%, lowest \deepseekcloud 24\%).  On average \appname identifies up to 2.5 times higher failure rates than the best SOTA approach per tested AUT.}
%     \label{fig:navi-ratio-results}
% \end{figure*}

For NaviQA, failure counts are shown in \autoref{fig:rq1-naviqa} (top). As in SafeQA, \appname consistently discovers more failures than both \rs and \gs across all AUTs. The highest failure counts are observed with \mistral, which is consistent with its smaller model size (7B). \appname also exhibits greater variability, indicating broader exploration. In \convnavitwo, \appname clearly outperforms the baselines in both absolute failures and failure ratios.

Failure rates are reported in \autoref{fig:rq1-naviqa} (bottom). Across all AUTs, failure ratios converge over time, confirming sufficient testing budget. With \appname, \gptfivechat achieves the lowest rate (24\%), followed by \deepseekcloud and \gptfouro, while \mistral reaches the highest failure rate (39\%). 
Overall comparison, \appname achieves, on average, a 2.5 times higher failure rate.
To assess statistical significance, we applied the Mann–Whitney U test~\cite{Wilcoxon1945} ($\alpha = 0.05$) and quantified effect size using Vargha–Delaney $A^{12}$~\cite{delaney}. The improvements of \appname over all baselines are statistically significant, with large effect sizes, for both case studies and the three AUTs.

\begin{tcolorbox}[boxrule=0pt,frame hidden,sharp corners,enhanced,borderline north={1pt}{0pt}{black},borderline south={1pt}{0pt}{black},boxsep=2pt,left=2pt,right=2pt,top=2.5pt,bottom=2pt]
\textbf{RQ\textsubscript{1} (effectiveness).} \appname detects significantly more failures and produces higher failure rates than randomized/combinatorial testing, and a state-of-the-art approach in safety testing of six LLMs. A second study with two AUT types and three LLM variants similarly demonstrates that \appname outperforms randomized/combinatorial testing in identifying failures.
\end{tcolorbox}

\subsubsection{Diversity (RQ\textsubscript{2})} 

\autoref{tab:diversity-results-coverage} presents the failure diversity results based on coverage. For SafeQA, \appname consistently surpasses \astral across all models. Although \rs and \gs occasionally achieve competitive coverage, these cases occur primarily on weaker AUTs, where failures are easier to trigger and diversify. As a result, their diversity scores are less indicative of true behavioral robustness. 
In contrast, on harder-to-test models such as \gptfivechat, where failure modes are significantly harder to expose (based on the results in \autoref{fig:rq1-safeqa}, \appname achieves the highest coverage (up to 98\%), demonstrating its strength. This highlights that \appname maintains diversity not through random dispersion, but through targeted exploration of challenging failure regions.
For both \convnavione and \convnavitwo, \appname matches or exceeds the coverage achieved by \rs and \gs across all AUTs.
In SafeQA, coverage differences between \appname and \astral are statistically significant, with large effect sizes favoring \appname. In NaviQA, no baseline shows a significant advantage over \appname. These findings align with prior observations that randomized methods may explore failures more broadly than optimization-based testing~\cite{HUMENIUK2023102990, Sorokin2024}.
However, \appname maintains competitive diversity while achieving substantially higher failure discovery (RQ\textsubscript{1}). Since \appname does not explicitly target diversity, exploring such strategies remains an open direction for future work.

\begin{tcolorbox}[boxrule=0pt,frame hidden,sharp corners,enhanced,borderline north={1pt}{0pt}{black},borderline south={1pt}{0pt}{black},boxsep=2pt,left=2pt,right=2pt,top=2.5pt,bottom=2pt]
    \textbf{RQ\textsubscript{2} (diversity).} 
    For the safety case study, \appname achieves consistently higher coverage than \astral. Although \rs and \gs occasionally match coverage, this occurs only on weaker AUTs. On stronger models (e.g., \gptfivechat) \appname obtains the highest diversity, confirming its effectiveness. For the navigational case study, \appname delivers higher coverage than the baselines.
    % For the safety case study, \appname achieves on average higher coverage results then \astral. For the majority of comparisons for safety \appname is surpassed by \gs and \rs. For the navigational case study it achieves in four out of six comparisons statistically significant higher entropy scores than \rs and \gs, with similar coverage results.
    % \appname achieves a comparable diversity in terms of failure coverage as \rs, \gs and the state of the art approach \astral. 
    % Specifically, for the navigational case study it achieves in four out of four comparison statistically significant higher entropy scores than \rs and \gs, with similar coverage results.
\end{tcolorbox}

\begin{table}[t]
\centering
\scriptsize
\caption{Results for RQ\textsubscript{2} (diversity). Coverage across failure clusters (\%, averaged over 10 runs).}
\label{tab:diversity-results-coverage}
\renewcommand{\arraystretch}{1.1}
\setlength{\tabcolsep}{4.5pt}
\begin{tabular}{llcccc}
\toprule
\textbf{Case Study} & \textbf{Model} & \textbf{\appname} & \textbf{\astral} & \textbf{\gs} & \textbf{\rs} \\
\midrule
\multirow{6}{*}{\textbf{SafeQA}} 
 & \deepseeklocal   & 79 & 56 & \textbf{89} & 89 \\
 & \mistral         & 82 & 58 & 90 & \textbf{92} \\
 & \deepseekcloud   & 79 & 65 & 88 & \textbf{89} \\
 & \gptfivechat     & \textbf{98} & 93 & 96 & 96 \\
 & \qwenthree       & 86 & 61 & \textbf{91} & 91 \\
 & \gptfouro        & \textbf{93} & 78 & 91 & 93 \\
\midrule
\multirow{3}{*}{\textbf{NaviQA-I}} 
 & \mistral         & \textbf{100} & -- & 100 & 100 \\
 & \deepseekcloud   & \textbf{100} & -- & 100 & 100 \\
 & \gptfivechat     & \textbf{96} & -- & 88 & 87 \\
\midrule
\multirow{1}{*}{\textbf{NaviQA-II}} 
 & \gptfivechat     & \textbf{99} & -- & 95 & 96 \\
\bottomrule
\end{tabular}
\end{table}

\begin{table*}[t]
\centering
\caption{Observed failure types in \convnavitwo with criticality and ratio of failures identified by \appname per type.}
\label{tab:system_failures}
% \resizebox{\textwidth}{!}{
\begin{tabular}{clp{5.5cm}p{4.2cm}cc}
% \begin{tabular}{clllcc}
\toprule
\textbf{ID} & \textbf{Failure Type} & \textbf{Failure Description} & \textbf{Example Answer} & \textbf{Criticality} & \textbf{Ratio (\%)} \\
\midrule
F1 & Endpoint Failure & AUT endpoint call fails. & --- & High & 91\\ 
F2 & Incorrect Rating & POI is returned but has an incorrect rating. & --- & Medium & 79 \\ 
F3 & Name Misinterpretation & Perturbation of the utterance causes the system to misinterpret a POI name constraint. & “I could not find ae supermarket.” & Medium & 83 \\ 
F4 & Language Misclassification & Perturbation of the utterance causes the system to incorrectly detect a language change. & “Schau me please a bar” $\rightarrow$ “Habe nichts passendes gefunden.” & High & 86 \\ 
F5 & Technical Output & System outputs technical information not intended for the user. & “You should apply the filter `payment:card'...” & High & 60 \\ 
F6 & Search Not Performed & System does not perform the search but repeats the request constraints. & “Let me search a German bar.” & Low & 48 \\ 
F7 & POI Retrieval & Requested POI exists but cannot be retrieved. & "Hmm, no Thai cafes nearby. Want me to look for Thai restaurants?" & High & 84 \\ 
F8 & Wrong Intent & The system misunderstands the intent of the request. & “I'm sorry, I cannot answer to that.” & High & 64 \\
F9 & Empty Output & System returns empty text. & --- & Medium & 57 \\
\bottomrule
\end{tabular}
% }
\end{table*}

\section{Threats To Validity}\label{sec:threats}

\head{Internal Validity} Internal validity concerns factors that might affect the correctness of our results. First, several analyses required thresholds (e.g., for clustering and distance metrics). Different threshold values might alter the results. However, we selected these based on preliminary experiments and default values, but other choices could lead to different outcomes. 
% \andrea{if that is not the case, remove this or, better, clarify that this does not happen}
Second, LLMs are inherently non-deterministic, producing different outputs for the same prompt. Additionally, API call durations can vary. We mitigated this by repeating runs and averaging results, but some residual randomness may persist. Finally, the LLM-based generator may occasionally produce utterances that do not exactly reflect the intended feature values, introducing noise in the test set. We evaluate the generator initially and monitor such cases. %, but they cannot be fully avoided. 

\head{Construct validity} Construct validity concerns the degree to which our measures and constructs reflect the phenomena of interest. First, we measure convergence via the ratio of discovered failures. Embedding distance checks help cluster related inputs, yet they cannot account for every possible variation. Second, in our case studies, we defined three custom judge dimensions with the help of domain experts and user data. While expert input improves realism, it also introduces subjectivity. We partially mitigated this by conducting correlation analysis with human evaluators, though replication with additional experts would further strengthen validity. In addition, we conducted a human evaluation to assess the accuracy of the LLM-generated test inputs. Finally, we relied on ASTRAL in an offline RAG configuration. While this setup ensures repeatability, it may not capture the performance of an OpenAI-based online RAG system.

\head{External Validity} External validity concerns the extent to which our results can be generalized. We studied two different case studies involving three AUTs and up to six LLMs. Although the systems and models differ in nature, the scope is still limited. Generalization to other domains, larger systems, or additional LLM families should be made with caution.

\section{Qualitative Evaluation}\label{sec:qualitative-analysis}

We performed a qualitative evaluation for the study \convnavitwo. In particular, we clustered using k-means all failures found by all approaches based on embeddings of system answers received. It should be noted, that this clustering differs from the analyses in $RQ_2$, as it is based on the system outputs rather than on test inputs. Then we sampled 100 failures per cluster and manually assigned each cluster to a specific failure type. In addition, we measured per cluster how many failures of each cluster are identified by \appname or by remaining approaches and reported all results in \autoref{tab:system_failures}.

% % \andrea{the ratio should add up to 100\%?}
% \autoref{tab:system_failures} summarizes the failure types manually identified in \convnavitwo.

% Each failure class is characterized by its behavioral manifestation by an illustrative example. 

% Further, we assessed its criticality as well as the proportion of failing inputs in which it was observed. %\ken{"each failure class..." This sentence is hard to understand} 
The identified failures range from endpoint and retrieval errors to misinterpretations caused by linguistic perturbations. We then interviewed a domain expert with 7 years of working experience in the testing of AI-enabled systems such as \convnavitwo and presented him the failures found and the types extracted. We then asked the following questions:

\textit{Q1: Are the failures consistent with the extracted failure types, and are the failure types realistic?} Following the expert, all the identified failure types are in line with the inspected failures, and can occur in \convnavitwo. 

\textit{Q2: Which of the failure types have already been witnessed when testing \convnavitwo before?} Based on the response, failure types F1, F2, F6, F7, F8, and F9 have already been frequently detected, F4 has been seen in different contexts but not yet in this expression (e.g., so far it only occurred when German street names were pronounced based on English pronunciation logic by an English native).
Only the failure types F3 and F5 have not been detected so far through testing. 
% \andrea{the question is about the time required to find failures, not about likelihood of occurrence}
% \lev{please check, is it better?}
% \andrea{do we have an idea if it takes engineers minutes, hours, or days of manual testing to find these failures? We need to show that the automated approach has a leverage}
% \lev{actually F3,F5 ocurr very seldon, but it does not mean it is difficult to find. need to rephrase}

\textit{Q3: What is the severity of the identified failures/types?} The majority of the identified failures are of high severity, highlighting the importance of preventing their propagation to the end user. Failure F6 is of low severity, as this type is difficult to avoid due to likely synchronization issues within the system.
% \textit{Q4. Is \appname a useful framework to be employed for testing conversational systems such as \convnavitwo?}
% Based on the expert, the so far unidentified failures F3 and F5 have criticality of medium/high making it especially relevant that tools such as \appname can help to identify such failures. We presented the expert how \appname is configurable in terms of feature dimensions, evaluation and failure criteria, The expert considers to leverage \appname for testing other functionalities of \convnavitwo such as for the car expert use case~\cite{rony-etal-2023-carexpert}.
Based on the interview, we can confirm that our testing uncovered a wide spectrum of distinct failure types. Especially, \appname achieved high detection rates in categories such as F3 and F5, which the domain expert identified as particularly challenging to reveal. This suggests that \appname is especially effective at uncovering uncommon failures that are likely to be missed by baseline methods. 

\section{Related work}\label{sec:related-work}

% \subsection{Testing of LLMs}
% \hl{Needs updates and adding papers and rewrite of below paragraphs (is taken over from Contextual paper).}
% To cope with the non-determinism and complexity of LLM applications, research has investigated approaches both to optimize prompts as well as to benchmark LLMs. EvoPrompt~\cite{evoprompt} and Protege~\cite{pryzant-etal-2023-automatic}, for instance, have been proposed to optimize the prompts using evolutionary algorithms. However, our scope is on identifying failing test inputs using an optimization heuristic and not refining prompts.

Several datasets have been developed to evaluate LLMs across different application domains~\cite{evoprompt,pryzant-etal-2023-automatic}. For safety testing, representative examples include BeaverTails~\cite{ji2023beavertails}, Do-Not-Answer~\cite{wang-etal-2024-answer}, and ToxiGen~\cite{hartvigsen-etal-2022-toxigen}. For holistic evaluation of general capabilities, benchmarks such as MMLU~\cite{wang2024mmlu} and BIG-Bench~\cite{bai2024mt} are commonly used. Furthermore, to assess conversational systems, multiple dialogue-oriented datasets have been introduced, including CoQA~\cite{reddy2019coqa}, MMDialog~\cite{feng2022mmdialog}, VACW~\cite{siegert-2020-alexa}, MultiWOZ 2.2~\cite{zang2020multiwoz}, and KVRET~\cite{eric2017keyvalue}, which particularly focus on tasks involving navigational requests and recommendations.
However, these datasets may already be included in the training data of the LLM under test, which undermines their effectiveness for evaluation. 

To address this challenge, several automated testing approaches have been proposed that generate variations or build upon existing datasets. In the following, we describe the main approaches and compare them with \appname.
For instance, Andriushchenko et al.~\cite{andriushchenko2025jailbreakingleadingsafetyalignedllms} tried to jailbreak LLMs by modifying a suffix that is appended to a prompt template, while the suffix content is selected based on a randomized search. They achieved up to 100\% attack rates on leading safety-aligned LLMs. However, this approach is tailored to robustness testing of LLMs, while \appname is generic, which also considers functional testing of LLM applications. 
METAL~\cite{metal} is a metamorphic testing framework that benchmarks LLMs across quality attributes such as robustness and fairness. However, it relies on the availability of an initial, diverse dataset where both inputs and expected outputs are already defined. Moreover, its testing process is static, as it does not incorporate a feedback loop to guide the generation of likely failing test cases. In contrast, \appname actively modifies test inputs to dynamically explore failure-inducing behaviors.

MORTAR~\cite{guo2024mortarmetamorphicmultiturntesting} is a framework to benchmark LLM-based dialogue systems by applying operations on the turn level, such as deletion, repetition, or shuffling of turns. Failures can be detected by evaluating outcomes and applying metamorphic relations. Also, this approach requires the existence of an initial dataset to compare the generated responses with.
Yoon et al.~\cite{ART2025juyeon} presented an adaptive randomized test selection approach that requires an initial dataset, but selects likely failure-revealing test inputs based on evaluations performed on existing data. This is done by computing the distance of a candidate test input to already generated test inputs, prioritizing test inputs with a higher distance. 
EvoTox~\cite{evotox} focuses on testing LLMs' toxicity, providing a systematic way to quantitatively surface toxic responses. ASTRAL~\cite{ASTRAL} is the most closely related automated test generation approach, and we include it in our empirical comparison. Similar to \appname, it discretizes the search domain and constructs a coverage matrix over feature combinations. However, ASTRAL faces two key limitations: it does not scale well to a high number of features, and it lacks a feedback loop from evaluated test inputs. In contrast, \appname incorporates an optimization process that leverages feedback to guide the generation of new, potentially failing test cases.

\section{Conclusion}\label{sec:conclusion}

In this paper, we introduced \appname, a search-based testing framework for benchmarking LLM-based applications. Through two case studies and three systems under test, covering both malicious input handling and task-oriented dialogues, we demonstrated that \appname consistently uncovers significantly more failures than randomization/combinatorial-based methods and a state-of-the-art automated testing baseline.
In our future work, we plan to extend \appname to applications involving multiple modalities, such as image and audio inputs, and to investigate its applicability to multi-turn dialogue scenarios, where interaction dynamics and conversational state introduce new challenges for testing.

\section{Data Availability}

The pipeline used to obtain the results discussed in this work and the results are available in our replication package~\cite{repo}.

\section*{Acknowledgements}
\addcontentsline{toc}{section}{Acknowledgements}
This research was funded by the Bavarian Ministry of Economic Affairs, Regional Development and Energy, and by the BMW Group. We thank all survey participants and colleagues from BMW who supported this work.

% \clearpage
% \balance
\bibliographystyle{IEEEtran}

\begin{thebibliography}{10}
\providecommand{\url}[1]{#1}
\csname url@samestyle\endcsname
\providecommand{\newblock}{\relax}
\providecommand{\bibinfo}[2]{#2}
\providecommand{\BIBentrySTDinterwordspacing}{\spaceskip=0pt\relax}
\providecommand{\BIBentryALTinterwordstretchfactor}{4}
\providecommand{\BIBentryALTinterwordspacing}{\spaceskip=\fontdimen2\font plus
\BIBentryALTinterwordstretchfactor\fontdimen3\font minus \fontdimen4\font\relax}
\providecommand{\BIBforeignlanguage}[2]{{%
\expandafter\ifx\csname l@#1\endcsname\relax
\typeout{** WARNING: IEEEtran.bst: No hyphenation pattern has been}%
\typeout{** loaded for the language `#1'. Using the pattern for}%
\typeout{** the default language instead.}%
\else
\language=\csname l@#1\endcsname
\fi
#2}}
\providecommand{\BIBdecl}{\relax}
\BIBdecl

\bibitem{zhang2025comprehensivesurveyprocessorientedautomatic}
Y.~Zhang, H.~Jin, D.~Meng, J.~Wang, and J.~Tan, ``A comprehensive survey on process-oriented automatic text summarization with exploration of llm-based methods,'' 2025.

\bibitem{elshin-etal-2024-general}
D.~Elshin, N.~Karpachev, B.~Gruzdev, I.~Golovanov, G.~Ivanov, A.~Antonov, N.~Skachkov, E.~Latypova, V.~Layner, E.~Enikeeva, D.~Popov, A.~Chekashev, V.~Negodin, V.~Frantsuzova, A.~Chernyshev, and K.~Denisov, ``From general {LLM} to translation: How we dramatically improve translation quality using human evaluation data for {LLM} finetuning,'' in \emph{Proceedings of the Ninth Conference on Machine Translation}, 2024.

\bibitem{duolingo2024max}
{Duolingo}. (2024) Duolingo max with gpt-4.

\bibitem{dong2025surveycodegenerationllmbased}
Y.~Dong, X.~Jiang, J.~Qian, T.~Wang, K.~Zhang, Z.~Jin, and G.~Li, ``A survey on code generation with llm-based agents,'' 2025.

\bibitem{jin2025llmsllmbasedagentssoftware}
H.~Jin, L.~Huang, H.~Cai, J.~Yan, B.~Li, and H.~Chen, ``From llms to llm-based agents for software engineering: A survey of current, challenges and future,'' 2025.

\bibitem{liu2024largelanguagemodelbasedagents}
J.~Liu, K.~Wang, Y.~Chen, X.~Peng, Z.~Chen, L.~Zhang, and Y.~Lou, ``Large language model-based agents for software engineering: A survey,'' 2024.

\bibitem{10.1145/3769082}
Z.~Sheng, Z.~Chen, S.~Gu, H.~Huang, G.~Gu, and J.~Huang, ``Llms in software security: A survey of vulnerability detection techniques and insights,'' \emph{ACM Comput. Surv.}, 2025.

\bibitem{friedl2023incarethinkingincarconversational}
K.~E. Friedl, A.~G. Khan, S.~R. Sahoo, M.~R. A.~H. Rony, J.~Germies, and C.~Süß, ``Inca: Rethinking in-car conversational system assessment leveraging large language models,'' 2023.

\bibitem{2025-Giebisch-IV}
R.~Giebisch, K.~E. Friedl, L.~Sorokin, and A.~Stocco, ``Automated factual benchmarking for in-car conversational systems using large language models,'' in \emph{Proceedings of the 36th IEEE Intelligent Vehicles Symposium}, ser. IV '25, 2025.

\bibitem{ahmed2025rag}
B.~S. Ahmed, L.~O. Baader, F.~Bayram, S.~Jagstedt, and P.~Magnusson, ``{Quality Assurance for LLM-RAG Systems: Empirical Insights from Tourism Application Testing},'' in \emph{2025 IEEE International Conference on Software Testing, Verification and Validation Workshops (ICSTW)}, 2025.

\bibitem{rapisarda2025qatesting}
R.~G. Rapisarda, D.~Ginelli, D.~Clerissi, and L.~Mariani, ``{ Test Case Generation for Dialogflow Task-Based Chatbots },'' in \emph{2025 IEEE International Conference on Software Testing, Verification and Validation Workshops (ICSTW)}, 2025.

\bibitem{bmw2024ces}
\BIBentryALTinterwordspacing
{BMW Group Press Release}. (2024) {Generative AI, Augmented Reality and Teleoperated Parking – The Digital Experience in the BMW of the Future at the Consumer Electronics Show (CES) 2024}. BMW Group. Accessed: 2025-10-17. [Online]. Available: \url{https://tinyurl.com/3cbapys8}
\BIBentrySTDinterwordspacing

\bibitem{2025-Guo-arxiv}
\BIBentryALTinterwordspacing
Y.~Gao, M.~Piccinini, Y.~Zhang, D.~Wang, K.~Moller, R.~Brusnicki, B.~Zarrouki, A.~Gambi, J.~F. Totz, K.~Storms, S.~Peters, A.~Stocco, B.~Alrifaee, M.~Pavone, and J.~Betz, ``Foundation models in autonomous driving: A survey on scenario generation and scenario analysis,'' 2025. [Online]. Available: \url{https://arxiv.org/abs/2506.11526}
\BIBentrySTDinterwordspacing

\bibitem{xie2025sorrybench}
T.~Xie, X.~Qi, Y.~Zeng, Y.~Huang, U.~M. Sehwag, K.~Huang, L.~He, B.~Wei, D.~Li, Y.~Sheng, R.~Jia, B.~Li, K.~Li, D.~Chen, P.~Henderson, and P.~Mittal, ``{SORRY}-bench: Systematically evaluating large language model safety refusal,'' in \emph{The Thirteenth International Conference on Learning Representations}, 2025.

\bibitem{zang2020multiwoz}
X.~Zang, A.~Rastogi, S.~Sunkara, R.~Gupta, J.~Zhang, and J.~Chen, ``Multiwoz 2.2: A dialogue dataset with additional annotation corrections and state tracking baselines,'' in \emph{Proceedings of the 2nd Workshop on Natural Language Processing for Conversational AI, ACL 2020}, 2020.

\bibitem{ji2023beavertails}
J.~Ji, M.~Liu, J.~Dai, X.~Pan, C.~Zhang, C.~Bian, C.~Zhang, R.~Sun, Y.~Wang, and Y.~Yang, ``Beavertails: Towards improved safety alignment of llm via a human-preference dataset,'' 2023.

\bibitem{bradbury2025combinatorial}
J.~S. Bradbury and R.~More, ``{ Addressing Data Leakage in HumanEval Using Combinatorial Test Design },'' in \emph{2025 IEEE Conference on Software Testing, Verification and Validation (ICST)}, 2025.

\bibitem{ASTRAL}
M.~Ugarte, P.~Valle, J.~A. Parejo, S.~Segura, and A.~Arrieta, ``Astral: A tool for the automated safety testing of large language models,'' in \emph{Proceedings of the 34th ACM SIGSOFT International Symposium on Software Testing and Analysis}, ser. ISSTA Companion '25, 2025.

\bibitem{Zeller17SBST}
A.~Zeller, ``Search-based testing and system testing: A marriage in heaven,'' ser. SBST, 2017.

\bibitem{5954405}
P.~McMinn, ``Search-based software testing: Past, present and future,'' in \emph{2011 IEEE Fourth International Conference on Software Testing, Verification and Validation Workshops}, 2011.

\bibitem{riccio2020model}
V.~Riccio and P.~Tonella, ``Model-based exploration of the frontier of behaviours for deep learning system testing,'' in \emph{Proceedings of the 28th ACM Joint Meeting on European Software Engineering Conference and Symposium on the Foundations of Software Engineering}, 2020.

\bibitem{2020-Riccio-EMSE}
V.~Riccio, G.~Jahangirova, A.~Stocco, N.~Humbatova, M.~Weiss, and P.~Tonella, ``{Testing Machine Learning based Systems: A Systematic Mapping},'' \emph{Empirical Software Engineering}, 2020.

\bibitem{2025-Weissl-TOSEM}
O.~Weißl, A.~Abdellatif, X.~Chen, G.~Merabishvili, V.~Riccio, S.~Kacianka, and A.~Stocco, ``Targeted deep learning system boundary testing,'' \emph{ACM Transactions on Software Engineering and Methodology}, 2025.

\bibitem{sbse2019ramirez}
A.~Ramírez, J.~R. Romero, and S.~Ventura, ``A survey of many-objective optimisation in search-based software engineering,'' \emph{Journal of Systems and Software}, vol. 149, 2019.

\bibitem{Abdessalem-ICSE18}
R.~{Ben Abdessalem}, S.~{Nejati}, L.~{C. Briand}, and T.~{Stifter}, ``Testing vision-based control systems using learnable evolutionary algorithms,'' in \emph{2018 IEEE/ACM 40th International Conference on Software Engineering (ICSE)}, 2018.

\bibitem{Abdessalem-ASE18-1}
R.~B. Abdessalem, A.~Panichella, S.~Nejati, L.~C. Briand, and T.~Stifter, ``Testing autonomous cars for feature interaction failures using many-objective search,'' in \emph{Proceedings of ASE '18}, ser. ASE 2018, 2018.

\bibitem{sorokin2023opensbt}
L.~Sorokin, T.~Munaro, D.~Safin, B.~H.-C. Liao, and A.~Molin, ``{OpenSBT}: A modular framework for search-based testing of automated driving systems,'' in \emph{Proceedings of the 2024 IEEE/ACM 46th International Conference on Software Engineering: Companion Proceedings}, ser. ICSE-Companion '24, 2024.

\bibitem{NejatiSSFMM23}
S.~Nejati, L.~Sorokin, D.~Safin, F.~Formica, M.~M. Mahboob, and C.~Menghi, ``Reflections on surrogate-assisted search-based testing: {A} taxonomy and two replication studies based on industrial {ADAS} and simulink models,'' \emph{Inf. Softw. Technol.}, vol. 163, 2023.

\bibitem{sorokin24svm}
L.~Sorokin and N.~Kerscher, ``Guiding the search towards failure-inducing test inputs using support vector machines,'' in \emph{Proceedings of the 5th IEEE/ACM International Workshop on Deep Learning for Testing and Testing for Deep Learning}, ser. DeepTest '24, 2024.

\bibitem{2025-Chen-EMSE}
X.~Chen, M.~Biagiola, V.~Riccio, M.~d'Amorim, and A.~Stocco, ``{XMutant: XAI-based Fuzzing for Deep Learning Systems},'' \emph{Empirical Software Engineering}, 2025.

\bibitem{2025-Maryam-ICST}
Maryam, M.~Biagiola, A.~Stocco, and V.~Riccio, ``Benchmarking generative ai models for deep learning test input generation,'' in \emph{Proceedings of the 18th IEEE International Conference on Software Testing, Verification and Validation}, ser. ICST '25.\hskip 1em plus 0.5em minus 0.4em\relax IEEE, 2025, p. 12 pages.

\bibitem{stocco2019misbehaviour}
A.~Stocco, M.~Weiss, M.~Calzana, and P.~Tonella, ``Misbehaviour prediction for autonomous driving systems,'' in \emph{Proceedings of the 42nd International Conference on Software Engineering}, ser. ICSE '20.\hskip 1em plus 0.5em minus 0.4em\relax ACM, jun 2020.

\bibitem{evoprompt}
Q.~Guo, R.~Wang, J.~Guo, B.~Li, K.~Song, X.~Tan, G.~Liu, J.~Bian, and Y.~Yang, ``Connecting large language models with evolutionary algorithms yields powerful prompt optimizers,'' in \emph{The Twelfth International Conference on Learning Representations}, 2024.

\bibitem{pryzant-etal-2023-automatic}
R.~Pryzant, D.~Iter, J.~Li, Y.~Lee, C.~Zhu, and M.~Zeng, ``Automatic prompt optimization with ``gradient descent'' and beam search,'' in \emph{Proceedings of the 2023 Conference on Empirical Methods in Natural Language Processing}, 2023.

\bibitem{repo}
L.~Sorokin and I.~Vasilev, ``Replication package,'' \url{https://github.com/ast-fortiss-tum/STELLAR}.

\bibitem{rony-etal-2023-carexpert}
M.~R. A.~H. Rony, C.~Suess, S.~R. Bhat, V.~Sudhi, J.~Schneider, M.~Vogel, R.~Teucher, K.~Friedl, and S.~Sahoo, ``{C}ar{E}xpert: Leveraging large language models for in-car conversational question answering,'' in \emph{Proceedings of the 2023 Conference on Empirical Methods in Natural Language Processing: Industry Track}, 2023.

\bibitem{9536732}
S.~Alharbi, M.~Alrazgan, A.~Alrashed, T.~Alnomasi, R.~Almojel, R.~Alharbi, S.~Alharbi, S.~Alturki, F.~Alshehri, and M.~Almojil, ``Automatic speech recognition: Systematic literature review,'' \emph{IEEE Access}, vol.~9, 2021.

\bibitem{lin-2004-rouge}
C.-Y. Lin, ``{ROUGE}: A package for automatic evaluation of summaries,'' in \emph{Text Summarization Branches Out}, 2004.

\bibitem{papineni2002bleu}
K.~Papineni, S.~Roukos, T.~Ward, and W.-J. Zhu, ``Bleu: a method for automatic evaluation of machine translation,'' in \emph{Proceedings of the 40th annual meeting on association for computational linguistics}.\hskip 1em plus 0.5em minus 0.4em\relax Association for Computational Linguistics, 2002.

\bibitem{Deb02NSGA2}
K.~Deb, A.~Pratap, S.~Agarwal, and T.~Meyarivan, ``A fast and elitist multiobjective genetic algorithm: Nsga-ii,'' \emph{IEEE Trans. Evol. Comput.}, vol.~6, no.~2, 2002.

\bibitem{deb2007adaptive}
K.~Deb, K.~Sindhya, and T.~Okabe, ``Self-adaptive simulated binary crossover for real-parameter optimization,'' in \emph{Proceedings of the 9th Annual Conference on Genetic and Evolutionary Computation}, ser. GECCO '07, 2007.

\bibitem{Mikolov:2013ohu}
T.~Mikolov, K.~Chen, G.~Corrado, and J.~Dean, ``{Efficient Estimation of Word Representations in Vector Space},'' 2013.

\bibitem{llamaindexsimilarity}
``Llamaindex,'' \url{https://developers.llamaindex.ai/python/framework-api-reference/evaluation/semantic_similarity/}.

\bibitem{Sorokin2024}
L.~Sorokin, D.~Safin, and S.~Nejati, ``Can search-based testing with pareto optimization effectively cover failure-revealing test inputs?'' \emph{Empirical Software Engineering}, vol.~30, no.~1, 2024.

\bibitem{munaro2026faultinjection}
T.~Munaro, M.~Turalija, S.~Barner, and M.~Halak, ``A systematic approach to fault injection test case generation in practice,'' in \emph{Software Engineering and Advanced Applications}, 2026.

\bibitem{Abdessalem2018TestingVC}
R.~B. Abdessalem, S.~Nejati, L.~C. Briand, and T.~Stifter, ``Testing vision-based control systems using learnable evolutionary algorithms,'' \emph{2018 IEEE/ACM 40th International Conference on Software Engineering (ICSE)}, 2018.

\bibitem{surrogate2024biagiola}
M.~Biagiola and P.~Tonella, ``Testing of deep reinforcement learning agents with surrogate models,'' \emph{ACM Trans. Softw. Eng. Methodol.}, vol.~33, no.~3, 2024.

\bibitem{feldt2016diversity}
R.~Feldt, S.~Poulding, D.~Clark, and S.~Yoo, ``Test set diameter: Quantifying the diversity of sets of test cases,'' in \emph{2016 IEEE International Conference on Software Testing, Verification and Validation (ICST)}, 2016.

\bibitem{wen-etal-2023-unveiling}
J.~Wen, P.~Ke, H.~Sun, Z.~Zhang, C.~Li, J.~Bai, and M.~Huang, ``Unveiling the implicit toxicity in large language models,'' in \emph{Proceedings of the 2023 Conference on Empirical Methods in Natural Language Processing}, 2023.

\bibitem{andriushchenko2025jailbreakingleadingsafetyalignedllms}
M.~Andriushchenko, F.~Croce, and N.~Flammarion, ``Jailbreaking leading safety-aligned llms with simple adaptive attacks,'' 2025.

\bibitem{Asghar16}
N.~Asghar, ``Yelp dataset challenge: Review rating prediction,'' \emph{CoRR}, vol. abs/1605.05362, 2016.

\bibitem{9260048}
K.~R. Shahapure and C.~Nicholas, ``Cluster quality analysis using silhouette score,'' in \emph{2020 IEEE 7th International Conference on Data Science and Advanced Analytics (DSAA)}, 2020.

\bibitem{metal}
S.~Hyun, M.~Guo, and M.~A. Babar, ``{ METAL: Metamorphic Testing Framework for Analyzing Large-Language Model Qualities },'' in \emph{2024 IEEE Conference on Software Testing, Verification and Validation (ICST)}, 2024.

\bibitem{pronouncing}
``Cmu-pronouncing library,'' 2022.

\bibitem{bortfeld2001fluent}
H.~Bortfeld, S.~Leon, J.~Bloom, M.~Schober, and S.~Brennan, ``Disfluency rates in conversation: Effects of age, relationship, topic, role, and gender,'' \emph{Language and speech}, vol.~44, 2001.

\bibitem{habicht2025benchmarkingcontextualunderstandingincar}
\BIBentryALTinterwordspacing
P.~Habicht, L.~Sorokin, A.~Saydemir, K.~E. Friedl, and A.~Stocco, ``Benchmarking contextual understanding for in-car conversational systems,'' 2025. [Online]. Available: \url{https://arxiv.org/abs/2512.12042}
\BIBentrySTDinterwordspacing

\bibitem{bai2024mt}
G.~Bai, J.~Liu, X.~Bu, Y.~He, J.~Liu, Z.~Zhou, Z.~Lin, W.~Su, T.~Ge, B.~Zheng \emph{et~al.}, ``Mt-bench-101: A fine-grained benchmark for evaluating large language models in multi-turn dialogues,'' \emph{arXiv preprint arXiv:2402.14762}, 2024.

\bibitem{Fleiss:1971}
J.~L. Fleiss, ``Measuring nominal scale agreement among many raters.'' \emph{Psychological bulletin}, vol.~76, no.~5, 1971.

\bibitem{cabrera1994logistic}
A.~Cabrera, ``Logistic regression analysis in higher education: An applied perspective,'' \emph{Higher education: Handbook of theory and research X/Agathon Press}, 1994.

\bibitem{Wilcoxon1945}
F.~Wilcoxon, ``Individual comparisons by ranking methods,'' \emph{Biometrics Bulletin}, vol.~1, no.~6, 1945.

\bibitem{delaney}
A.~Vargha and H.~D. Delaney, ``A critique and improvement of the cl common language effect size statistics of mcgraw and wong,'' \emph{Journal of Educational and Behavioral Statistics}, vol.~25, no.~2, 2000.

\bibitem{HUMENIUK2023102990}
D.~Humeniuk, F.~Khomh, and G.~Antoniol, ``Ambiegen: A search-based framework for autonomous systems testingimage 1,'' \emph{Science of Computer Programming}, vol. 230, 2023.

\bibitem{wang-etal-2024-answer}
Y.~Wang, H.~Li, X.~Han, P.~Nakov, and T.~Baldwin, ``Do-not-answer: Evaluating safeguards in {LLM}s,'' in \emph{Findings of the Association for Computational Linguistics: EACL 2024}, 2024.

\bibitem{hartvigsen-etal-2022-toxigen}
T.~Hartvigsen, S.~Gabriel, H.~Palangi, M.~Sap, D.~Ray, and E.~Kamar, ``{T}oxi{G}en: A large-scale machine-generated dataset for adversarial and implicit hate speech detection,'' in \emph{Proceedings of the 60th Annual Meeting of the Association for Computational Linguistics (Volume 1: Long Papers)}, 2022.

\bibitem{wang2024mmlu}
Y.~Wang, X.~Ma, G.~Zhang, Y.~Ni, A.~Chandra, S.~Guo, W.~Ren, A.~Arulraj, X.~He, Z.~Jiang \emph{et~al.}, ``Mmlu-pro: A more robust and challenging multi-task language understanding benchmark,'' \emph{arXiv preprint arXiv:2406.01574}, 2024.

\bibitem{reddy2019coqa}
S.~Reddy, D.~Chen, and C.~D. Manning, ``Coqa: A conversational question answering challenge,'' \emph{Transactions of the Association for Computational Linguistics}, vol.~7, 2019.

\bibitem{feng2022mmdialog}
J.~Feng, Q.~Sun, C.~Xu, P.~Zhao, Y.~Yang, C.~Tao, D.~Zhao, and Q.~Lin, ``Mmdialog: A large-scale multi-turn dialogue dataset towards multi-modal open-domain conversation,'' \emph{arXiv preprint arXiv:2211.05719}, 2022.

\bibitem{siegert-2020-alexa}
I.~Siegert, ``\BIBforeignlanguage{eng}{``{A}lexa in the wild'' {--} collecting unconstrained conversations with a modern voice assistant in a public environment},'' in \emph{\BIBforeignlanguage{eng}{Proceedings of the Twelfth Language Resources and Evaluation Conference}}, 2020.

\bibitem{eric2017keyvalue}
M.~Eric, L.~Krishnan, F.~Charette, and C.~D. Manning, ``Key-value retrieval networks for task-oriented dialogue,'' in \emph{Proceedings of the 18th Annual SIGdial Meeting on Discourse and Dialogue (SIGDIAL)}, 2017.

\bibitem{guo2024mortarmetamorphicmultiturntesting}
G.~Guo, A.~Aleti, N.~Neelofar, and C.~Tantithamthavorn, ``Mortar: Metamorphic multi-turn testing for llm-based dialogue systems,'' 2024.

\bibitem{ART2025juyeon}
J.~Yoon, R.~Feldt, and S.~Yoo, ``{ Adaptive Testing for LLM-Based Applications: A Diversity-Based Approach },'' in \emph{2025 IEEE International Conference on Software Testing, Verification and Validation Workshops (ICSTW)}, 2025.

\bibitem{evotox}
S.~Corbo, L.~Bancale, V.~D. Gennaro, L.~Lestingi, V.~Scotti, and M.~Camilli, ``How toxic can you get? search-based toxicity testing for large language models,'' \emph{IEEE Transactions on Software Engineering}, vol.~51, no.~11, pp. 3056--3071, 2025.

\end{thebibliography}

% \input{9-appendix}

\end{document}